\documentclass[groupedaddress,superscriptaddress,aps,showpacs,amsmath,amssymb,floatfix, twocolumn,pra,a4]{revtex4-1}

 \usepackage{graphicx,float}
 \usepackage{subfigure}
 \usepackage{dcolumn}
\usepackage{bm}
\usepackage{mathrsfs}
\usepackage{txfonts}
\usepackage{CJK}
\usepackage{SIunits}
\usepackage{epsfig}
\usepackage{lipsum}
\usepackage{textcomp}
\usepackage{color}

 \newenvironment{SChinese}{%
  \CJKfamily{gbsn}%
 \CJKtilde
  \CJKnospace}{}

\allowdisplaybreaks[2]

\begin{document}

\begin{CJK}{UTF8}{} 
\begin{SChinese}

\title{Tunable slowing, storing and releasing of a weak microwave field}

 \author{Keyu Xia (夏可宇)}  %
 \email{keyu.xia@mq.edu.au}
 \affiliation{Centre for Engineered Quantum Systems, Department of Physics and Astronomy, Macquarie University, NSW 2109, Australia}
%

\date{\today}

\begin{abstract}
We study the slowing, storing and releasing of microwave pulses in a superconducting circuits composed of two coplanar waveguide resonators and a superconducting transmon-type qubit. The quantum interference analogy to electromagnetically induced transparency is created in two coupled resonators. By tuning the resonance frequency of the transmon, we dynamically tune the effective coupling between the resonators. Via the modulation of the coupling, we show the tunable true time delay of microwave pulses at the single-photon level. We also store the microwave field in a high-Q resonator and release the signal from it to the output port. Our scheme promises applications in both quantum information processing and classical wireless communications.
    
\end{abstract}

\pacs{42.50.Gy, 41.20.Jb, 85.25.Cp, 42.60.Da}

\maketitle

\end{SChinese}
 \end{CJK}

\section{Introduction}

Since the electromagnetically induced transparency (EIT) was discovered as a quantum interference in atoms \cite{AEIT1}, it has been widely used to enhance nonlinear susceptibility \cite{AEIT2,Kerr}, store \cite{AStore1} and slow light \cite{AEIT2} in an atomic medium. 
The light trapped in coupled resonators display the pathway interference \cite{OEIT1,OEIT2,OEIT3,OEIT4, OCavity1} similar to the quantum interference in atoms \cite{AEIT2}. This path interference induces the optical analogue of EIT in coupled optical resonators \cite{OEIT1,OEIT2,OEIT3} and optomechanics \cite{OEMIT1,OEMIT2} has been demonstrated for slow light.

Although the optical analogue of EIT and its application for slowing light has been well studied, the tunable delay of a purely microwave pulse is still a challenging \cite{MWPhoton}. On the other hand, a tunable control over true time delay of microwave signals is essential for the applications in radar and satellite communications \cite{MWPhoton}. A design to indirectly delay the microwave signals modulates the optical signals by the microwave signal and then slow down the group velocity of the optical signal. In this way, the microwave signal added to an optical carrier is delayed \cite{MWPhoton}.

It is attractive for the wireless communications to slow down a purely microwave signal on a microscopic chip. Using the metamaterial analog of EIT, a microwave pulse is delayed by $40\%$ of the pulse width by two coupled metamaterial slabs \cite{MWDelay3}. In this configuration, the transparency window and the time delay are fixed once the setup is fabricated. Very recently, the tunable delay of microwave pulses has been demonstrated by using the microwave analog of EIT in electromechanics \cite{MWDelay1,MWDelay2}. The storage of microwave electromganetic waves has also been realized in the matamaterial analogue of EIT \cite{StoreMWinEIT}. However, the performance of the EIT-based slowing of single-photon pulses in the quantum regime is unclear.
The microwave field can also be stored in and retrieved from an electron spin ensemble \cite{MWStorage1,QMemory} or an electromechanical resonator \cite{MWTransfer}. It is essential to delay a single microwave photon for the superconducting-circuit-based quantum information processing. Yin et al. has reported the storage and retrieve of quantum microwave photons by dynamically tuning the coupling of a resonator to an open transmission line \cite{MWStorage2}.
Moreover, the slowing microwave fields using the Aulter-Townes effects in ``artificial'' atoms is in progress \cite{AT1,AT2,AT3}.


In this paper, We extend the basic idea directly controlling the photon-photon interaction between cavities in our previous work \cite{Idea} to the superconducting circuits. We propose a scheme to directly tune the coupling between two coplanar waveguide (CPW) resonators and subsequently to create the tunable microwave analogue of EIT. This tunable EIT allows one to control the group delay in microwave pulse propagation. By switching on-off the coupling between two resonators, we can also store a microwave field in a high-Q resonator and then release the field to the output port on-demand. Using the single-photon scattering model, we show the time delay of the microwave propagation at the single-photon level. The propagation of single microwave photons in real space is presented as a numerical proof of our scheme working in the quantum regime.

The paper is organized as follows. In Sec. II we introduce the setups of the systems and the relevant model for a microwave analogue of EIT. In Sec. III we then introduce the single-photon scattering model for numerical simulations of storing and slowing of microwave photons. The results are presented in Sec. IV and a conclusion of this work is given in Sec. V.

\section{System and Model}
In our previous work \cite{Idea}, we proposed a method to directly control the photon-photon interaction between two optical cavities using a $\Lambda-$type system. Here we apply the idea to control the coupling between two CPW resonators via tuning the transition frequency of a superconducting transmon qubit. 

\begin{figure*}
 \includegraphics[width=0.48\linewidth]{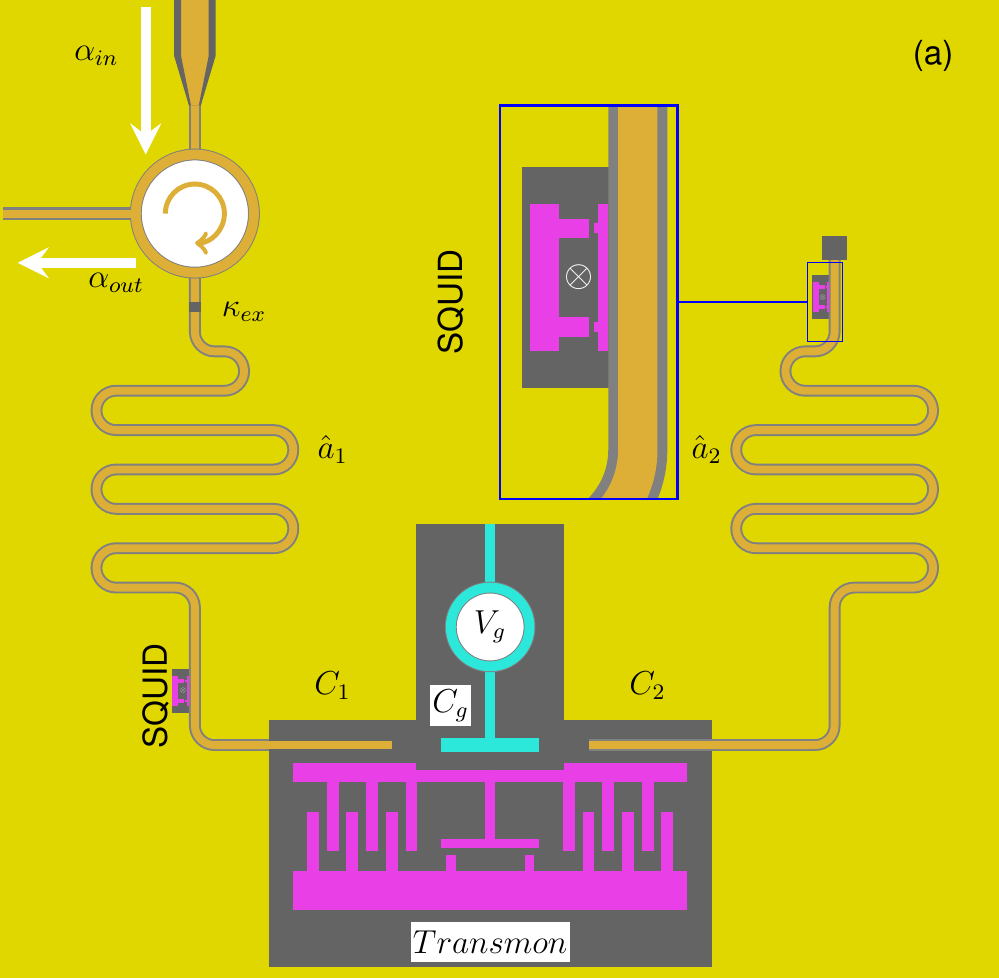}
 \includegraphics[width=0.48\linewidth]{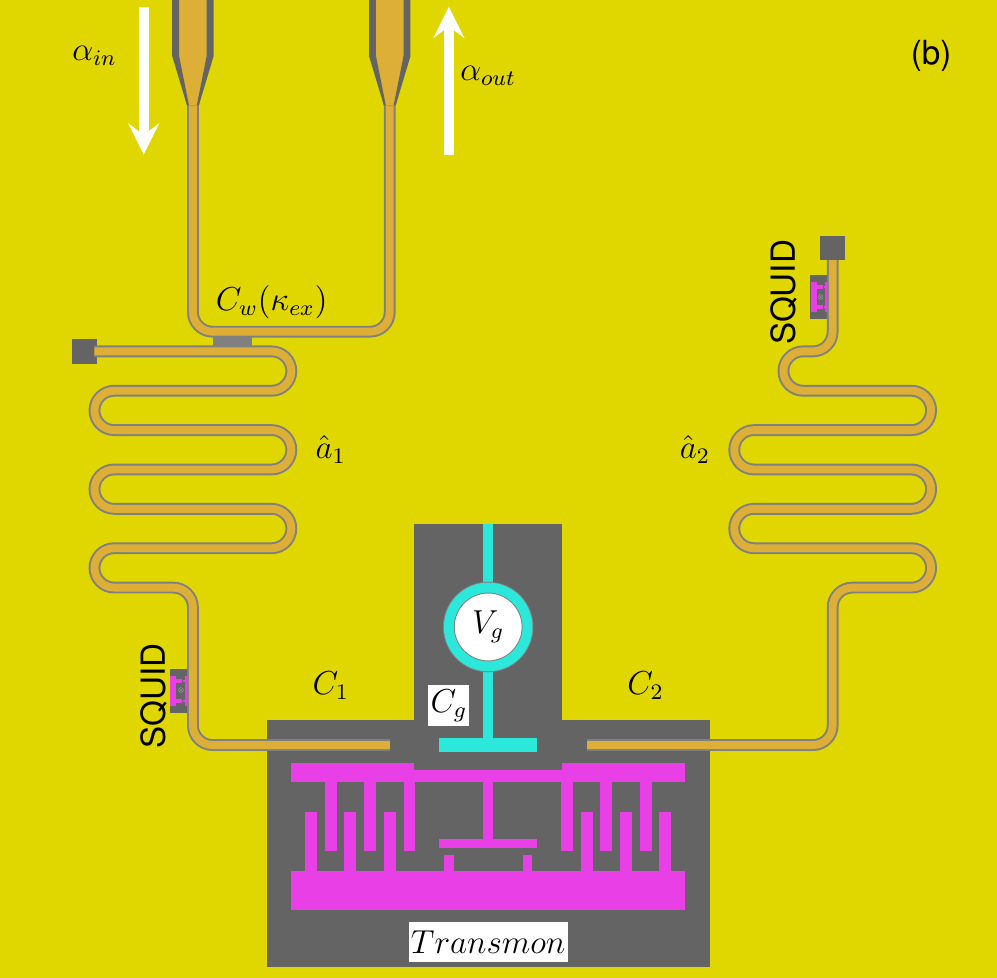} \\
\caption{(Color online) (a) direct(end)-coupling setup for storing and releasing of and (b) side-coupling setup for slowing of a weak microwave field. Two CPW resonator modes $\hat{a}_1$ and $\hat{a}_2$ couples to a superconducting transmon via their mutual capacitors $C_1$ and $C_2$. A gate voltage $V_g$ is used to bias the transmon via a gate capacitor $C_g$. The SQUIDs biased by magnetic fluxes are used to compensate the AC Stark shift induced by the transmon.}\label{fig:system}
\end{figure*}

Our system is shown in Fig.~\ref{fig:system}. Its structure is similar to the design by Schoelkopf et al. \cite{Design,ExpSetup}, but the transmon qubit with the excited state $|e\rangle$ and the ground state $|g\rangle$ simultaneously couples to two separate CPW resonators via their mutual capacitors $C_1$ and $C_2$, respectively. Besides, the first resonator capacitively couples to the transmission line yielding an external coupling $\kappa_{ex}$. While the second resonator only couples to the transmon. The incoming and outgoing microwave fields travel in the transmission line.
The gate voltage $V_g$ is used to bias the qubit and dynamically tune the transition frequency $\omega_q$ of qubit. The transition frequency of the transmon can also be tuned by a biased flux \cite{TransmonFluxBias}.

The first setup shown in Fig.~\ref{fig:system}(a) is used for slowing, storing and releasing of microwave signals. A microwave pulse is incident into the resonator 1 and then reflected to the ouput port. The reflected pulse is isolated by a circulator from the input port. To do storage, first, the effective coupling $h_e$ is on and the resonator mode $\hat{a}_2$ is excited. Then $h_e$ is turned off and the photon is stored in the high-Q resonator 2. To retrieve the photon, we turn on $h_e$ again. The photon stored in the resonator 2 excites the resonator mode $\hat{a}_1$ and subsequently goes toward the output port.
The second design shown in Fig.~\ref{fig:system}(b) is only used to slow down the microwave signals traveling in an open transmission line because the photons retrieved from the resonator 2 can enter both the input and output ports. So it can not be used to store and then release a photon to the targeted output port. 

In both setup, we apply superconducting quantum interference devices (SQUIDs) to tune the resonance frequency $\omega_{r_j}$ of the $j$the CPW resonator \cite{Tunefr1} and their frequency difference $\delta=\omega_{r_2} - \omega_{r_1}$, with $j\in \{1,2\}$. The resonators can also be tuned by a magnetic field with the help of a built-in SQUID \cite{Tunefr2,Tunefr3}. Detailedly, the SQUIDs dispersively couple with a strength $g_{s_j}$ to the nearby CPW resonator via their mutual capacitors. We assume that the SQUIDs can be modeled as a two-level system with the excited and ground states $|e_{s_j}\rangle$ and $|g_{s_j}\rangle$. The detuning between the $j$th SQUID and the $j$th resonator modes are $\Delta_{s_j}= \omega_{s_j} - \omega_{r_j}$ where $\omega_{s_j}$ denotes the transition frequency of the $j$th SQUID. In the non-resonant (dispersive) regime, the SQUIDs induce a shift equal to $-|g_{s_j}|^2 /\Delta_{s_j}$ in the resonance frequency of resonators \cite{Tunefr1}. This small frequency shift is used to cancel the AC Stark shift due to the coupling to the transmon. We apply a magnetic flux $\Phi_{s_j}$  to tune the transition frequency $\omega_{s_j}$ and subsequently the detuning $\Delta_{s_j}$. In doing so, we can control the effective resonance frequencies of the resonators. In the following investigation, we simply assume the resonance frequencies of the resonators are tunable but neglect the process to change them.

\begin{figure}
 \includegraphics[width=0.8\linewidth]{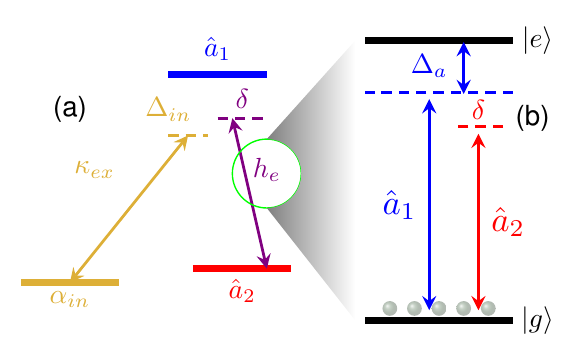} \\
\caption{(Color online) (a) Level-diagram picture, showing three "Levels" that represent the optical modes $\hat{a}_1$, $\hat{a}_2$ and the "probe" of optical waveguide mode $\alpha_{in}$. (b) Level diagram, showing how to tune the effective coupling $h_e$ via a two-level superconducting transmon qubit simultaneously coupling to two microwave resonators. The small balls above the level $|g\rangle$ indicates the population.}\label{fig:model}
\end{figure}

The quantum interference in our system in Fig.~\ref{fig:system} can be understood in a $\Lambda-$type three "Level" diagram shown in Fig.~\ref{fig:model}(a).    
In the dressed mode picture, by analogy to the dressed state view of EIT \cite{AEIT2}, it becomes clear that the absorption of the external driving by the two dressed modes composing of $\hat{a}_1$ and $\hat{a}_2$ is canceled, and the coupling $h_e$ can be used to switch the system from absorptive to transmittive/reflective in a narrow band around cavity resonance. In this case, the effective coupling between two cavity modes $\hat{a}_1$ and $\hat{a}_2$ creates an EIT-like transparency window for the external driving $\alpha_{in}$. The transmission profile depends on the resonance frequencies of modes $\hat{a}_1$ and $\hat{a}_2$, the frequency of driving $\alpha_{in}$, the coupling $h_e$ and the decay rates of two resonators.

Now we turn to explain the idea how to tune the coupling $h_e$ between two CPW resonators. As shown in Fig.~\ref{fig:model}(b), two cavity modes $\hat{a}_1$ and $\hat{a}_2$ dispersively couples to the transmon denoting the two-level's transition frequency and associated quantum eigenstates as $\omega_q$ and $|j\rangle$, $j\in\{e,g\}$. For simplicity, we assume that two modes couple identically to the transmon at a rate $g$, and $|\delta| \ll |\Delta_a|$. Both quantum fields $\hat{a}_1$ and $\hat{a}_2$ induce Stark shifts on levels $|g\rangle$ and $|e\rangle$. The value of this shift $\Delta_{Stark} = -|g|^2 \langle (\hat{a}_1^\dag + \hat{a}_2^\dag )(\hat{a}_1 + \hat{a}_2 )\rangle /\Delta_a$. In the situation of $|\Delta_a|^2 \gg |g|^2 \langle \hat{a}_1^\dag \hat{a}_1 \rangle, |g|^2 \langle \hat{a}_2^\dag \hat{a}_2 \rangle$, the excitation of the state $|e\rangle$ is negligible, i.e. $\langle \hat\sigma_{ee} \rangle \sim 0$, where we set $\hat{\sigma}_{ij}=|i\rangle \langle j|$. This Stark shift effectively yields a coherent interaction $h_e (\hat{a}_1^\dag \hat{a}_2 + \hat{a}_2^\dag \hat{a}_1)$ between the two resonator modes with a strength 
\begin{equation}\label{eq:He}
 h_e = -|g|^2 \langle \hat{\sigma}_{gg} \rangle /\Delta_a \sim -|g|^2 /\Delta_a \,.
\end{equation}
From Eq. (\ref{eq:He}), we see the manner to tune the effective coupling between two resonators is to tune the transition frequency $\omega_q$ of the transmon and subsequently the detuning $\Delta_a$.

Applying the Eq. (\ref{eq:He}) to the model, the effective Hamiltonian after eliminating the transmon takes the form ($\hbar=1$) (Appendix A)
\begin{equation}\label{eq:EffH}
\begin{split}
 \hat{H} = & \Delta'_{in} \hat{a}_1^\dag \hat{a}_1 +(\Delta'_{in}+\delta)\hat{a}_2^\dag \hat{a}_2  + 
h_e (\hat{a}_1^\dag \hat{a}_2 + \hat{a}_2^\dag \hat{a}_1) \\
 & + i\sqrt{2\kappa_{ex}} (\alpha_{in} \hat{a}_1^\dag - \alpha_{in}^* \hat{a}_1) \,,
\end{split}
\end{equation}
with $\Delta'_{in}=\Delta_{in} - |g|^2/\Delta_a$.
Then time-evolution of the system can be obtained by solving the Langevin equations
\begin{equation}\label{eq:LEq}
\begin{split}
 \frac{\partial \hat{a}_1}{\partial t} &= -(i\Delta'_{in} + \kappa'_1)  \hat{a}_1 - i h_e \hat{a}_2 + \sqrt{2\kappa_{ex}}\alpha_{in} \,,\\
 \frac{\partial \hat{a}_2}{\partial t} &= -(i\Delta'_{in} + i\delta + \kappa_2) \hat{a}_2 - i h_e \hat{a}_1 \,,
\end{split}
\end{equation}
where the decay rate $\kappa'_1$ of resonator 1 includes two contributions: $\kappa'_1 =\kappa_{1} + \kappa_{ex}$. $\kappa_1$ is the intrinsic decay rate and $\kappa_{ex}$ denotes the loss rate due to the external coupling to the resonator. While $\kappa_2$ only arises from the intrinsic loss of the resonator 2.
In the case of a weak coherent driving, the amplitude of intracavity fields can be approximated using the quantum average value of operators $\langle \hat{a}_1\rangle =\alpha$ and $\langle \hat{a}_2\rangle=\beta$, respectively. 
Using the input-output relation \cite{InputOutput1,InputOutput2}, the output field transmitted through or reflected by the transmon is given in terms of the input and intracavity fields as 
\begin{equation}\label{eq:InputOutput}
 \alpha_{out}= -\alpha_{in} + \sqrt{2\kappa_{ex}}\alpha \,.
\end{equation}
Immediately, we have the output in the steady state
\begin{equation}\label{eq:sss}
 \alpha_{ss}= -\alpha_{in} + \frac{2\kappa_{ex}[\kappa_2 + i(\delta + \Delta'_{in})] \alpha_{in}}{h_e^2 + (\kappa'_1 + i \Delta'_{in})[\kappa_2 + i(\delta + \Delta'_{in})]} \,.
\end{equation}

The transmission amplitude is defined here as $t=\alpha_{out}/\alpha_{in}$. The power transmission \textemdash the ratio of the power transmitted through the transmon divided by the input power \textemdash is calculated as $\mathcal{T}=|t|^2$. The coupling $h_e$ between two CPW resonators does not only induce a strong modulation of the transmission of the input field, at the same time it causes a rapid phase dispersion $\phi(\Delta'_{in}) = arg(t)$ leading to a "group delay" $\tau_g$ given by
\begin{equation}\label{eq:Gdelay}
 \tau_g = \frac{d \phi}{d\Delta'_{in}} \,,
\end{equation}
around the transparency window. Therefore, the tunability of the coupling $h_e$ allows one to control the group delay $\tau_g$. For identical cavities and on-resonance driving, i.e. $\Delta_{in}=0,\delta=0$ the group delay is given by
\begin{equation}\label{eq:delay}
 \tau_g = 2\kappa_{ex}\frac{h_e^2 -\kappa_2^2}{(h_e^2 + \kappa'_1\kappa_2)^2} \,.
\end{equation}
It is limited by the decay rate $\kappa_2$ of resonator 2. The maximum available group delay for $h_e^2 \gg \kappa_2^2, \kappa'_1\kappa_2$ is $\tau_g = 2\kappa_{ex}/h_e^2$.

Clearly, the steady-state solution of the Langevin equation Eq.~(\ref{eq:LEq}) can provide the EIT-like transmission or reflection spectrum. The solution is also useful to estimate the group delay $\tau_g$.

\section{Single-photon scattering model}
Besides the transmission or reflection in the steady state, the propagation in a real space is also interesting. In this section, we provide the single-photon scattering model developed by Fan et al. \cite{OL30p2001,PRA79p023837,PRA79p023838} for the study of the slowing of a microwave single-photon pulse in the following section.

We have discussed the manner to tune the photon-photon interaction between resonators \cite{Idea}. Here we directly consider an tunable effective coupling $h_e$ but neglect how to modulate $h_e$ by tuning the transition frequency of the transmon. In this case, our reduced system only consists of the open transmission line and two CPW resonators with their tunable coupling $h_e$.

For slowing light using the setup in Fig. \ref{fig:system}(b), the transmission line supports two counter-propagating modes: the incoming waveguide mode, $\hat{c}^\dag_T(x)$, from the input port and the other waveguide mode, $\hat{c}^\dag_R(x)$, reflected by the CPW resonator 1. Both two traveling modes interact with the CPW resonator 1 coupling to the second resonator mediated by the transmon. $\hat{c}^\dag_T(x)$ and $\hat{c}^\dag_R(x)$ create the photon at $x$ traveling from and to the input port, respectively. Since we will be interested in a narrow bandwidth single-photon pulses with a central frequency $\omega_{in}$ corresponding to a wave vector $k_0$, we can linearize the dispersion of the transmission line in the vicinity of $\omega_{in}$. 
 After the linearization, the effective Hamiltonian of the system we study then takes the form \cite{OL30p2001,PRA79p023837,PRA79p023838} after an energy shift of $\omega_{in}$
\begin{equation} \label{eq:Heff}
\begin{split}
 \hat{H}_{eff} =& -iv_g\int dx  \hat{c}_T(x)^\dag \frac{\partial}{\partial x}  \hat{c}_T(x) \\
& +iv_g\int dx \hat{c}_R(x)^\dag \frac{\partial}{\partial x}  \hat{c}_R(x)  \\
& + (\Delta'_{in} -i\kappa_1)\hat{a}_1^\dag \hat{a}_1 + (\Delta'_{in} +\delta -i\kappa_2)\hat{a}_2^\dag \hat{a}_2 \\
& + \int dx V\delta(x) (\hat{c}_T^\dag(x)\hat{a}_1  +\hat{a}_1^\dag \hat{c}_T(x)) \\
& + \int dx V\delta(x) (\hat{c}_R^\dag(x)\hat{a}_1  +\hat{a}_1^\dag \hat{c}_R(x)) \\
& + h_e (\hat{a}_1^\dag\hat{a}_2  +\hat{a}_2^\dag \hat{c}_1) \,,
\end{split}
\end{equation}
where $v_g$ is the group velocity of the traveling photon around $k_0$ in the transmission line. $V$ is the coupling strength between the transmission line and the resonator 1. Note that $V^2/v_g = \kappa_{ex}$ is the decay rate of resonator due to the external coupling to the transmission line and has the unit of frequency \cite{OL30p2001,PRA79p023837,PRA79p023838}.
In general, an arbitrary single-photon state $|\Phi(t)\rangle$ can be expressed as \cite{OL30p2001,PRA79p023837,PRA79p023838} 
\begin{equation}\label{eq:GState}
 \begin{split}
  |\Phi(t)\rangle = & \int dx [\tilde{\phi}_T(x,t)\hat{c}^\dag_T(x)  + \tilde{\phi}_R(x,t)\hat{c}^\dag_R(x)] |\varnothing\rangle \\
 & + \tilde{e}_1 \hat{a}^\dag_1 |\varnothing\rangle + \tilde{e}_2 \hat{a}^\dag_2 |\varnothing\rangle \,,
 \end{split}
\end{equation}
where $|\varnothing\rangle$ is the vacuum, which has zero photon and has the atom in the ground state. $\tilde{\phi}_{T/R}(x,t)$ is the single-photon wave function in the $T/R$ mode at the position $x$ and the time $t$.  $\tilde{e}_{1/2}$ is the excitation amplitude of the CPW resonator modes. We consider the propagation of the single-photon wave packets which can be derived from the Schr\"{o}dinger equation
\begin{equation} \label{eq:SDG}
 i\hbar \frac{\partial |\Phi(t)\rangle}{\partial t} = H_{eff} |\Phi(t)\rangle \,.
\end{equation}
Substitution of Eqs. ~(\ref{eq:EffH}) and ~(\ref{eq:GState}) into Eq.~(\ref{eq:SDG}) gives the following set of equations of motion in the real space
\begin{subequations}
\begin{align}\label{eq:Delay}
 \frac{\partial \tilde{\phi}_T(x,t)}{\partial t}  = &-v_g \frac{\partial \tilde{\phi}_T(x,t)}{\partial x} - iV\delta(x)\tilde{e}_1(t) \,, \\
 \frac{\partial \tilde{\phi}_R(x,t)}{\partial t} = &v_g \frac{\partial \tilde{\phi}_R(x,t)}{\partial x} - iV\delta(x)\tilde{e}_1(t) \,,\\
 \frac{\partial\tilde{e}_1(t)}{\partial t} = &-i(\Delta'_{in} -i\kappa_1)\tilde{e}_1 (t) -ih_e(t)\tilde{e}_2(t) \\ \nonumber
 & - iV\delta(x)\tilde{\phi}_{T}(x,t) -iV\delta(x)\tilde{\phi}_{R}(x,t) \,, \\
 \frac{\partial\tilde{e}_2(t)}{\partial t}  = & -i(\Delta'_{in} +\delta -i\kappa_2)\tilde{e}_2(t) -ih_e(t)\tilde{e}_1(t) \,.
\end{align}
\end{subequations}
For any given initial state $|\Phi(t)\rangle$, the dynamics of the system can be obtained directly by integrating this set of equations. The dynamics of our system is different from those of the previous works \cite{PRA79p023838,PRA82p063839}. In our system, the first resonator mode $\hat{a}_1$ couples both traveling modes $\hat{c}_T$ and $\hat{c}_R$, while the second resonator mode $\hat{a}_2$ only interacts with the mode $\hat{a}_1$. Furthermore, the intermode interaction $h_e$ can be tuned by the transmon.

For storage using the setup in Fig.~\ref{fig:system}(a), the microwave photons travels in the same transmission line before or after interaction with the resonator. The incoming pulse can be reflected by the transmon back to the transmission line. The microwave photon stored in the resonator can be also released to this branch of the transmission line. Following Shen and Fan \cite{PRA79p023837}, we define a field $\hat{c}_T^\dag(x)$ for a traveling mode in the transmission line such that $\hat{c}_T^\dag(x<x_0)$ describes an incoming photon that is moving toward the resonator at $x$, and $\hat{c}_T^\dag(x>x_0)$ describes an outgoing photon leaving the resonator at $2x_0 -x$. $x_0$ is the position of the resonator. In order to take into account the phase shift $\phi$ occurring during the reflection at the end of the waveguide, we write the Hamiltonian as
\begin{equation} \label{eq:Heff}
\begin{split}
 \hat{H}_{eff} =& -iv_g\int dx  \hat{c}_T(x)^\dag \frac{\partial}{\partial x}  \hat{c}_T(x) \,, \\
& -\int dx v_g \phi \frac{\partial f(x)}{\partial x}\hat{c}_T(x)^\dag \hat{c}_T(x) \,, \\
& + (\Delta'_{in} -i\kappa_1)\hat{a}_1^\dag \hat{a}_1 + (\Delta'_{in} +\delta -i\kappa_2)\hat{a}_2^\dag \hat{a}_2 \,,\\
& + \int dx V\delta(x) (\hat{c}_T^\dag(x)\hat{a}_1  +\hat{a}_1^\dag \hat{c}_T(x)) \,, \\
& + h_e (\hat{a}_1^\dag\hat{a}_2  +\hat{a}_2^\dag \hat{c}_1) \,,
\end{split}
\end{equation}
where $f(x)$ is a switch-on function with the general property that $lim_{x\rightarrow -\infty} f(x)=0$ and $lim_{x\rightarrow +\infty} f(x)=1$ in a short spatial extent. For computation purpose, we take $f(x)=\frac{1}{1+e^{-(x-x_0)/f_a}}$ with $f_a=0.5$ spatial step, otherwise the specific form of $f(x)$ is unimportant.
Consideration of a general single-excitation state 
\begin{equation}
  |\Phi(t)\rangle =  \int dx \tilde{\phi}(x,t)\hat{c}_T^\dag(x) |\varnothing\rangle 
 + \tilde{e}_1 \hat{a}^\dag_1 |\varnothing\rangle + \tilde{e}_2 \hat{a}^\dag_2 |\varnothing\rangle \,
\end{equation}
yields the equation of motion
\begin{subequations}
\begin{align}\label{eq:Store}
 \frac{\partial \tilde{\phi}(x,t)}{\partial t}  = &-v_g \frac{\partial \tilde{\phi}(x,t)}{\partial x} - iV\delta(x)\tilde{e}_1(t) \\ \nonumber
&  +iv_g\phi \frac{\partial f}{\partial x} \tilde{\phi}(x,t) \,,  \\
 \frac{\partial\tilde{e}_1(t)}{\partial t} = &-i(\Delta'_{in} -i\kappa_1)\tilde{e}_1 (t) -ih_e(t)\tilde{e}_2(t) \\ \nonumber
 & - iV\delta(x)\tilde{\phi}(x,t) \,, \\
 \frac{\partial\tilde{e}_2(t)}{\partial t}  = & -i(\Delta'_{in} +\delta -i\kappa_2)\tilde{e}_2(t) -ih_e(t)\tilde{e}_1(t) \,. 
\end{align}
\end{subequations}

The direct (end)-coupling model in Fig. 1(a) is essentially different from the side-coupling model in Fig. 1(b) \cite{PRA79p023837}. Therefore, the Hamiltonian Eq. (12) (Eq.(8)) and the differential equations Eq. (14) (Eq.(11)) only describes the interaction and propagation of microwave fields in the setup of Fig. 1(a) (1(b)).

\section{Results}
We use the Langevin equation to study the transmission and reflection spectrum of the system. Then we present the numerical validation of the capability of our system for slowing, storing and releasing of single-photon wave packets in the quantum regime. 

Throughout the following investigation, we apply a critical coupling $\kappa_{ex}=\kappa_1$ and assume that the decay rate of the resonator 2 is negligible, i.e. $\kappa_2=0$. This is practice if the resonator 1 has a low-Q with $Q_1\sim 10^3$ \cite{ExpSetup} and the resonator 2 is optimized to be high-Q, $Q_2 \sim 10^6$ \cite{Tunefr1}. For simplicity, we assume two identical resonators such that $\delta=0$ and $g_1=g_2=g$. A similar setup usable for our goals has been fabricated in experiment \cite{ExpSetup}. In our numerical simulation of the single-photon scattering, we assume that $\kappa_1=2\pi \times 5$~\mega\hertz. Thus, a coupling of $g=2\pi \times 100$~\mega\hertz~\cite{ExpSetup} can yield $h_e=2\kappa_1$ for $\Delta_1=10g$ which is the maximum in our simulation.

\subsection{Steady-state Solution}

\begin{figure}
 \includegraphics[width=0.48\linewidth]{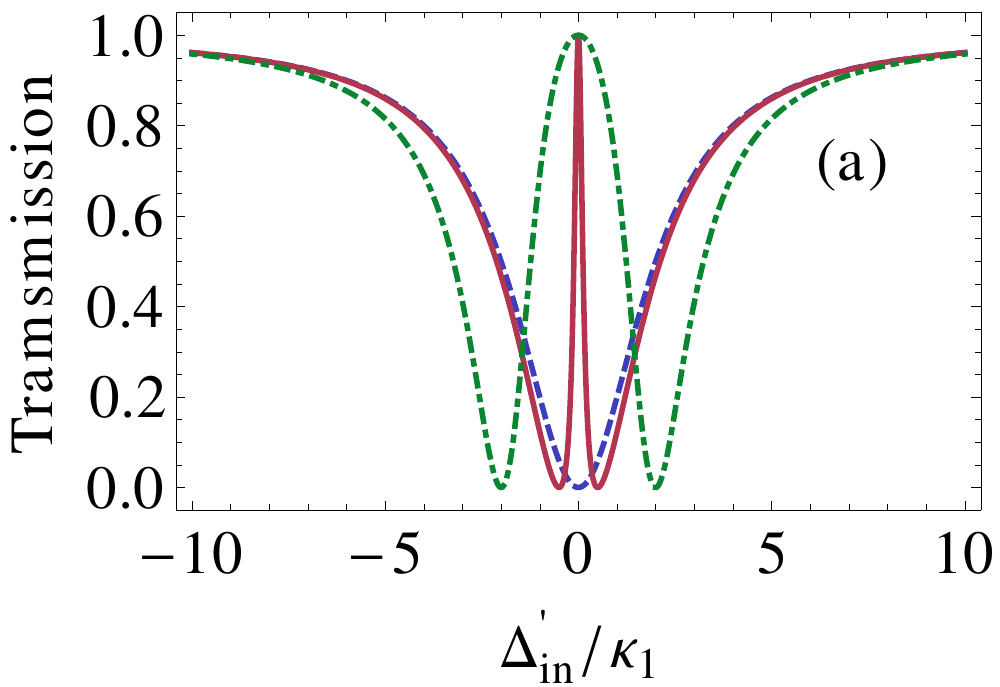}
 \includegraphics[width=0.48\linewidth]{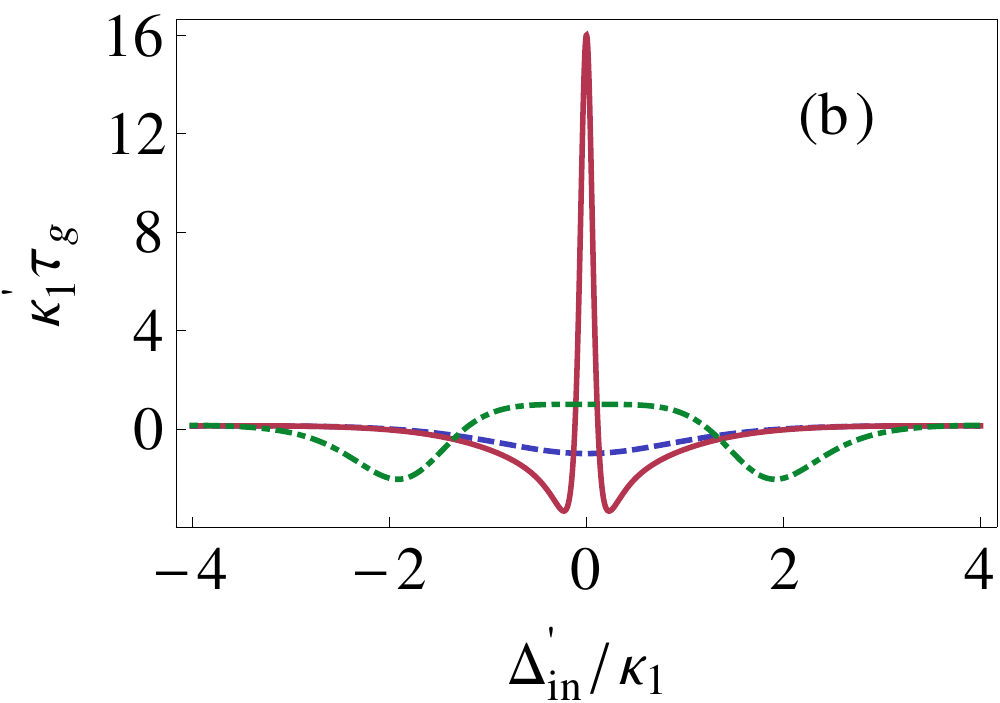}\\
\caption{(Color online) (a) Transmission $\mathcal{T}$ of the incoming waveguide mode and (b) the group delay of the transmitted microwave pules for different effective couplings $h_e/\kappa'_1=0,0.25,1$, respectively. The dashed blue lines for $h_e=0$, red solid lines for $h_e=0.25\kappa'_1$ and green dot-dashed lines for $h_e=\kappa'_1$.}\label{fig:sss}
\end{figure}

When an array of resonators couple to each other, the analogue of EIT occurs due to the pathway interference \cite{OEIT1,OEIT2,OEIT3,OEIT4, OCavity1}. For our system here, the microwave field in the CPW resonators 1 capacitively couples to the traveling field in the transmission line and interacts with the mode in the CPW resonator 2 mediated by the transmon. As a result, the reflection spectrum in Fig. ~\ref{fig:system}(a) and the transmission spectrum in Fig. ~\ref{fig:system}(b) display EIT profiles (see Fig.~\ref{fig:sss}(a)). This microwave analogue of EIT can be used to slow microwave pulses. Both of the transmission/reflection spectrum and the group delay (see Fig.~\ref{fig:sss}) can be calculated by solving the Langevin equation (\ref{eq:LEq}).

If the coupling between resonators is static, the group delay is fixed. Our system provides a manner to dynamically tune the effective coupling between the resonators. We tune the transition frequency of the transmon and subsequently the detuning $\Delta_a$ between the transmon and two resonators. Thus, the effective coupling $h_e=-|g|^2/\Delta_a$ can be modulated in time. 
In Fig.~\ref{fig:sss} the transmission/reflection spectrum and the group delay are compared for different couplings $h_e$. For $h_e=0$, no EIT-like profile appear. The microwave field is absorbed in the first setup (Fig.~\ref{fig:system}(a)) or reflected in the second setup (Fig.~\ref{fig:system}(b)) by the resonator 1. When $h_e>\kappa_2$, the EIT-like profile appear. For example, for $h_e=0.25\kappa'_1$, a narrow transparent window opens and the group delay can be $\kappa'_1\tau_g=16$, as shown in Fig.~\ref{fig:sss}(b). The group delay decreases quickly as the coupling $h_e$ increases. For $h_e>\kappa'_1$, the EIT-like window becomes broader but the delay of microwave pulses is very small. So, we use a small coupling $0<h_e<\kappa'_1$ for the slowing microwave photons.

\subsection{Single-photon Scattering}

Although the EIT-based slowing and storing of coherent microwave pulses has been demonstrated \cite{MWDelay2,MWDelay3}, it is unclear how well the EIT works for the slowing and storing of a microwave signal in the quantum regime. Here we study the storing and then releasing, and slowing of a single microwave photon by solving the motion of the single-photon scattering model. The initial input is a Gaussian pulse $\tilde{\phi}(x,0)=\sqrt[4]{\tau^2/\pi} e^{-(x-x_0)^2/2\tau^2}$ where $\tau$ is the spatial duration of pulse. The input is normalized to yield a single excitation, $\int_{-\infty}^{+\infty} \tilde{\phi}^*(x,0)\tilde{\phi}(x,0) dx=1$.

\begin{figure}
 \includegraphics[width=0.48\linewidth]{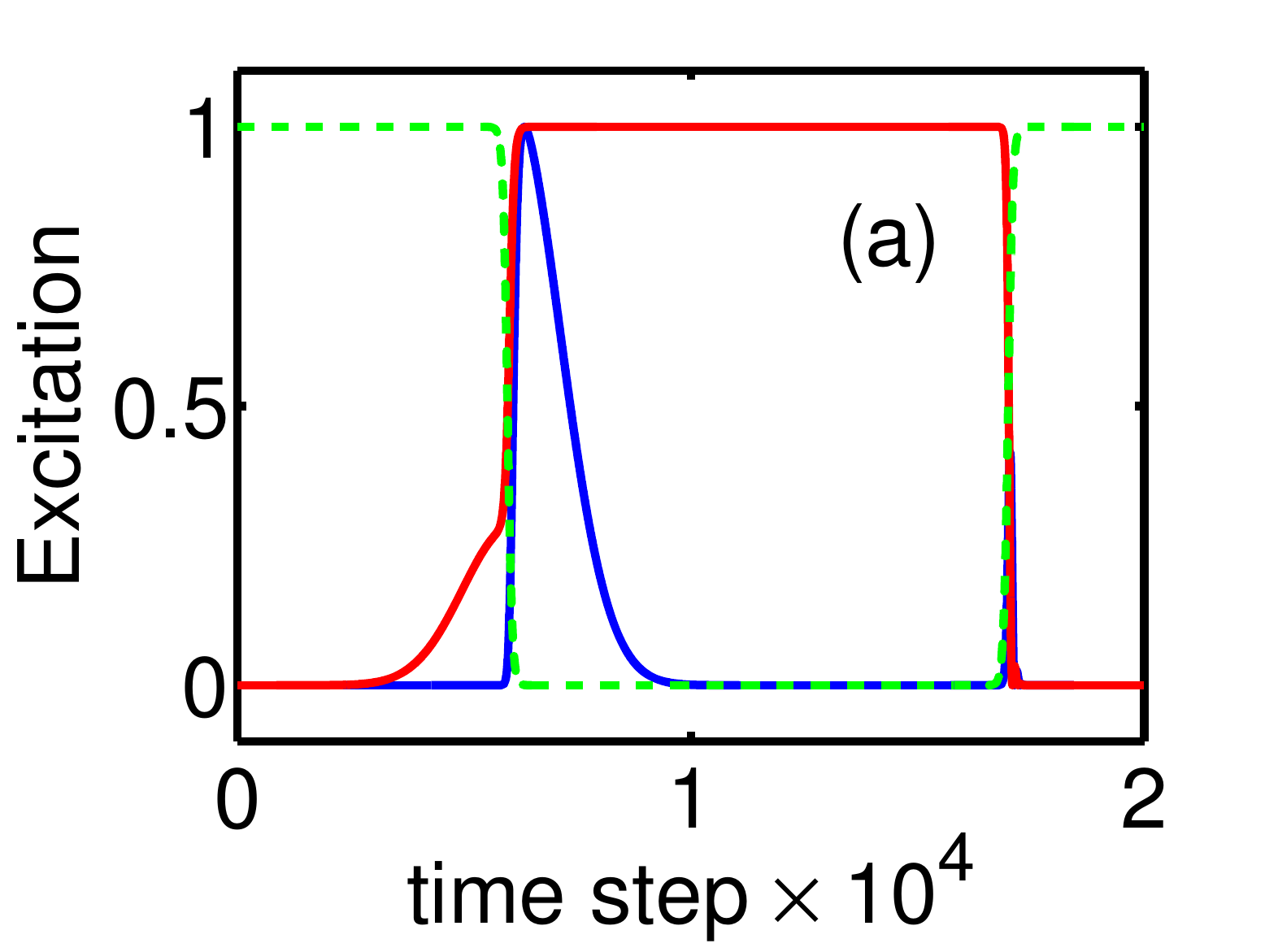}
 \includegraphics[width=0.48\linewidth]{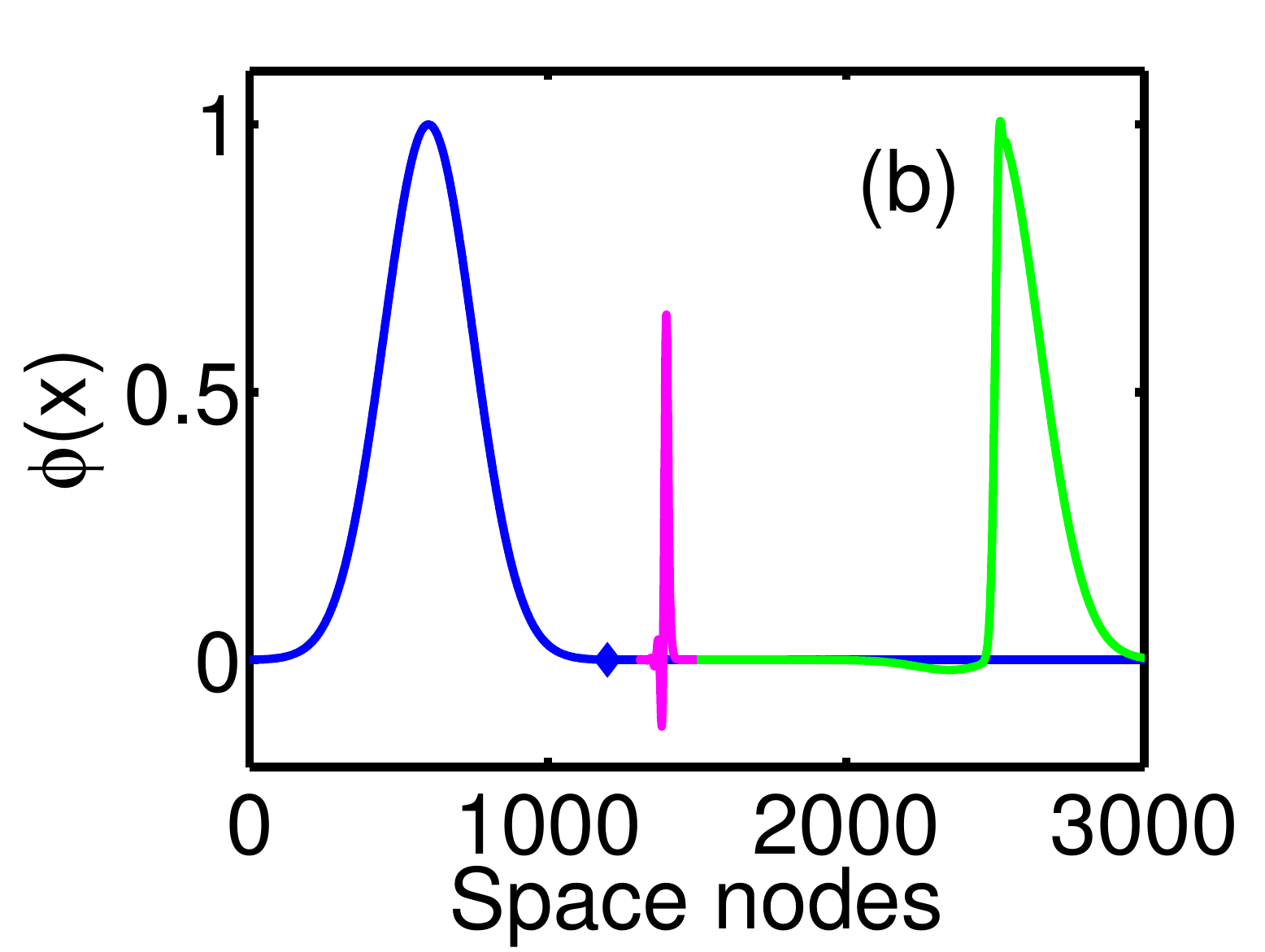}\\
\caption{(Color online) Storing and releasing of microwave single-photon pulses by solving Eq. (14). (a) the normalized time-evolution of excitation $|\tilde{e}_{1,2}|^2$ of the CPW resonators and the time-dependent effective coupling $h_e(t)$ (green dashed line) with the maximum $max(h_e(t)) =2\kappa_1$. The blue solid line shows the excitation of the resonator mode $\hat{a}_1$, red solid line for the mode $\hat{a}_2$. Excitations are normalized by their maximal numbers. (b) the propagation of single-photon pulses showing the storage and retrieve of microwave pulses. The wave functions are normalized. The blue line is for the input pulse, green line for the transmitted factor of pulse and magenta line shows the retrieved pulse. $h_e=2\kappa_1$. The diamond indicates the position of the resonator at $x_0=1200$.}\label{fig:Store}
\end{figure}

In Fig.~\ref{fig:Store}, we first tune on the effective coupling, $h_e=2\kappa_1$. The input pulse excites the resonator modes. Due to the transparent EIT window, a large fraction of the pulse is reflected by the resonator. Then the coupling is switched off within a short time. The excitation in the resonator 2 is stored for a time $\tau_s$ corresponding to a space delay $v_g\tau_s$. This fraction of the excitation is released to the transmission line again through the resonator 1 when the coupling turns on. It can be clearly seen that a retrieved pulse appears (magenta line) after the reflected one (green line). Because there is only a fraction of excitation can be stored and retrieved, the EIT-based scheme is useful for the storage of a coherent state.

\begin{figure}
 \includegraphics[width=0.48\linewidth]{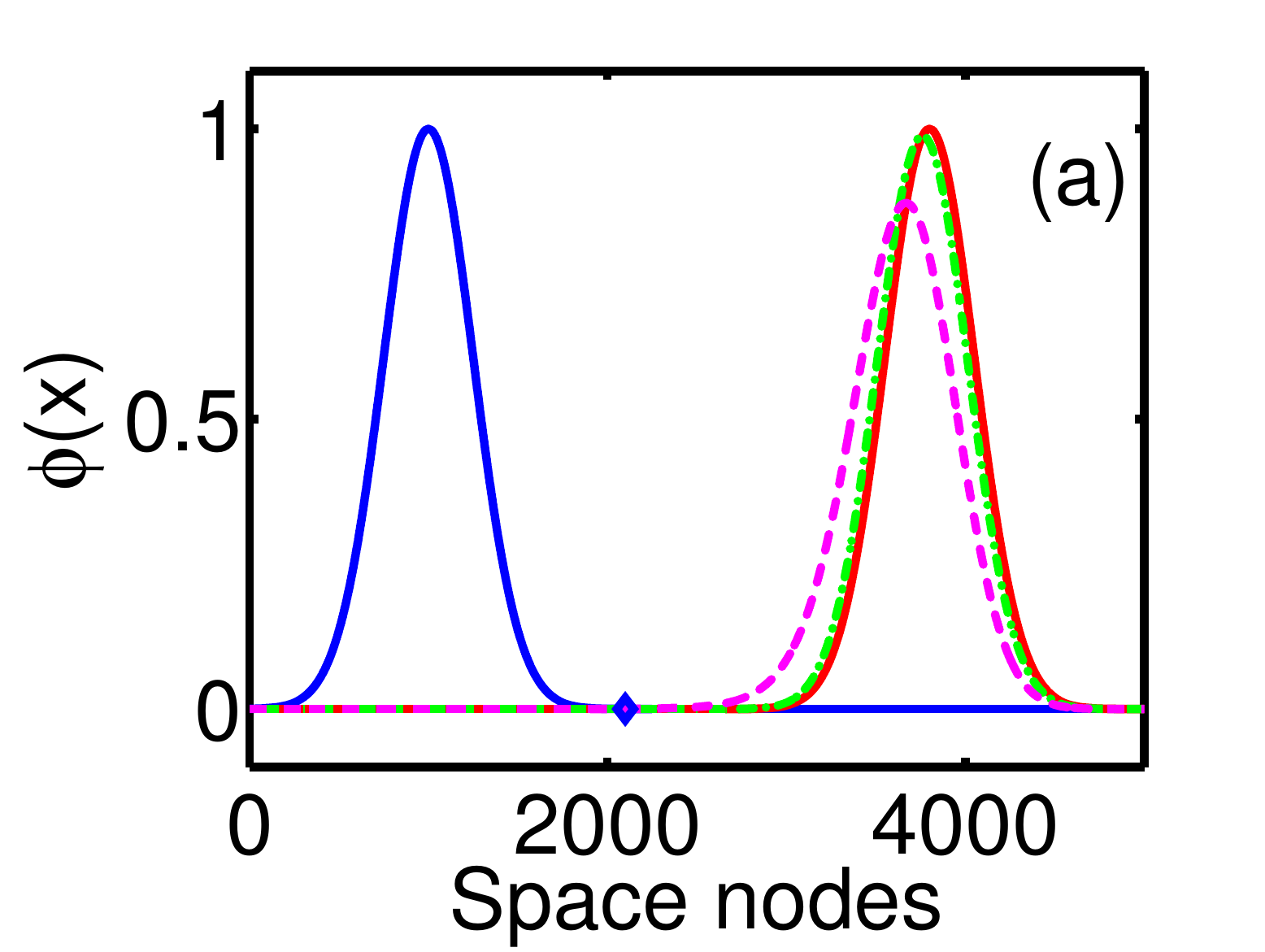} 
\includegraphics[width=0.48\linewidth]{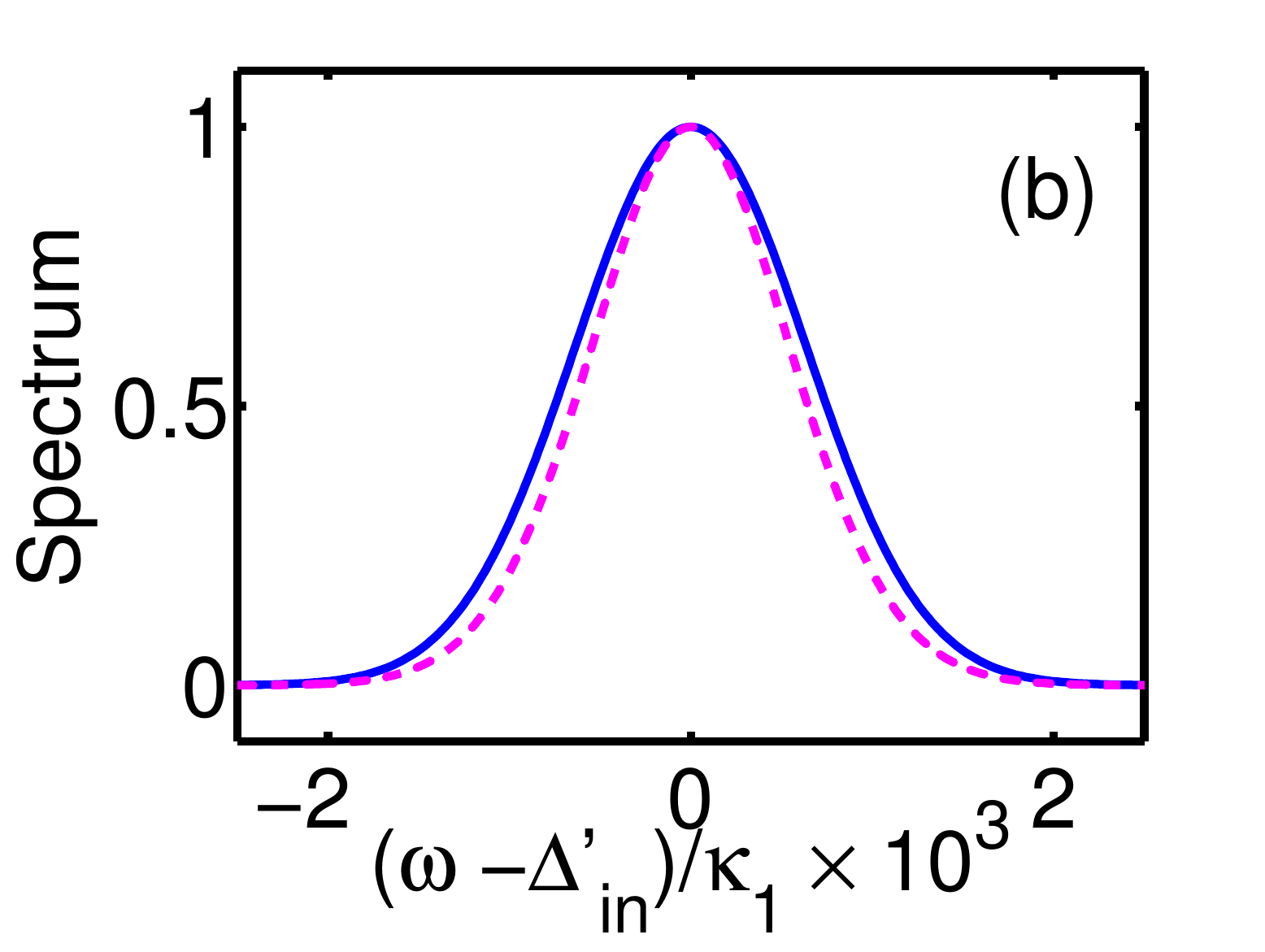}\\
\caption{(Color online) Slowing of microwave pulses at the single-photon level for various couplings $h_e$ by solving Eq. (11). (a) Propagation of the wave functions. All wave functions are normalized. The blue line is the input pulse. Red line for the free propagation in the absence of the resonators. Green dot-dashed line for $h_e=0.5\kappa_1$ and magenta dashed line for $h_e=0.25\kappa_1$. The diamond indicates the position of the resonator at $x_0=2100$. (b) Normalized spectrum of the input wave function (blue line) and the slowing one for $h_e=0.25\kappa_1$ (magenta dashed line).}\label{fig:Delay}
\end{figure}

Now we go on to the slowing single-photon pulses. Both setups in Fig.~\ref{fig:system} have the same capability to slow down the propagation of microwave pulses, shown in Fig.~\ref{fig:Delay}. The group delay is crucially dependent on the effective coupling $h_e$. For $h_e\geqslant \kappa_1$, the group delay is very small (green line). While $h_e \ll \kappa_1$ promises a considerable group delay. For example, $\tau_g$ can be larger than $50\% \tau$ if $h_e=0.25\kappa_1$. Although the slowing wave function has a smaller amplitude in comparison with the input pulse in space, the wave function broadens slightly in space. This can be seen from Fig.~\ref{fig:Delay}(b) because the transmitted pulse has a narrower spectrum. As a result, more than $86\%$ of excitation are reserved in the delayed pulse. 
Thus a cascade of two device can provide a group delay longer than the duration of pulse.
 
\section{Conclusion}
We proposed a scheme to directly control the coupling between two CPW resonators using a superconducting transmon-type qubit. This tunable interaction between microwave photons allows us to turn on/off the microwave analogue of the EIT. Using the single-photon scattering model, our simulations showed that we can tune the group delay of single-photon  microwave pulses. We also can store and release a microwave pulse. Our scheme provides a manner for the slowing and storing of the microwave photons in both of the microwave-based quantum information processing and the classical wireless communications.

\section*{Acknowledgements}
This work is supported by the ARC via the Centre of Excellence in Engineered Quantum Systems (EQuS), project number CE110001013.

\renewcommand{\theequation}{A\arabic{equation}}
\setcounter{equation}{0}
\section*{Appendix A}\label{App:A}
Here we provide the detail regarding the intermediate steps to obtain Eq. (\ref{eq:EffH}) in the main text. The Hamiltonian of the full system including the transmon takes the form
\begin{equation}
 \begin{split}
  \hat{H} = & \frac{\omega_q}{2}\sigma_z + \sum_{j=1,2} \omega_{r_j}\hat{a}_j^\dag \hat{a}_j\\
      & + i\sqrt{2\kappa_{ex}} \left ( \alpha_{in}e^{-i\omega_{in} t} \hat{a}_1^\dag - \alpha_{in}^*e^{i\omega_{in} t} \hat{a}_1\right) \\
    & + \sum_{j=1,2} \left(g_j^* \hat{a}_j^\dag \sigma_- + g_j \sigma_+ \hat{a}_j \right)
 \end{split}
\end{equation}
In the frame defined by the unitary transformation 
\begin{equation}
 \hat{U} = exp\left\{-i\left( \omega_{in}/2 \sigma_z + \sum_{j=1,2} \omega_{in}\hat{a}_j^\dag \hat{a}_j \right) t \right\}\,,
\end{equation}
the Hamiltonian becomes
\begin{equation} \label{eq:ApproxH}
 \begin{split}
  \hat{H} = & \frac{\Delta_a + \Delta_{in}}{2}\sigma_z +  \Delta_{in}\hat{a}_1^\dag \hat{a}_1+  (\Delta_{in} + \delta)\hat{a}_1^\dag \hat{a}_1\\
      & + i\sqrt{2\kappa_{ex}} \left ( \alpha_{in} \hat{a}_1^\dag - \alpha_{in}^* \hat{a}_1\right) \\
    & + \sum_{j=1,2} \left(g_j^* \hat{a}_j^\dag \sigma_- + g_j \sigma_+ \hat{a}_j \right) \,,
 \end{split}
\end{equation}
where the detunings are defined as $\Delta_a=\omega_q - \omega_{r_1}$, $\Delta_{in}=\omega_{r_1}-\omega_{in}$ and $\delta = \omega_{r_2} - \omega_{r_1}$. 
To use the projection-operator formalism \cite{Raman1} to derive the effective Hamiltonian, we divide the Hamiltonian Eq. (\ref{eq:ApproxH}) into
\begin{equation}\nonumber
 \hat{H} = \hat{H_e} + \hat{H_g} + \hat{V}_- +  \hat{V}_+ + \hat{H}_r \,,
\end{equation}
with
\begin{align}\nonumber
 \hat{H}_e = & \frac{\Delta_a + \Delta_{in}}{2} |e\rangle \langle e| \,,  \\ \nonumber
 \hat{H}_g = & -\frac{\Delta_a + \Delta_{in}}{2} |g\rangle \langle g|\,,  \\ \nonumber
 \hat{H}_r = & \Delta_{in}\hat{a}_1^\dag \hat{a}_1+  (\Delta_{in} + \delta)\hat{a}_1^\dag \hat{a}_1 \\ \nonumber
      & + i\sqrt{2\kappa_{ex}} \left ( \alpha_{in} \hat{a}_1^\dag - \alpha_{in}^* \hat{a}_1\right) \,,  \\ \nonumber
 \hat{V}_+ = & \sum_{j=1,2}\hat{V}_{+,j} \,,  \\ \nonumber
 \hat{V}_- = & \hat{V}_+^\dag \,,  \\ \nonumber
\end{align}
where $\hat{H}_r$ is independent of the state of the transmon. The perturbative excitation $\hat{V}_{+,j} = g_j \sigma_+ \hat{a}_j$ connects the transmon and the $j$th resonator.
Assuming that $|\Delta_a|$ is much larger than $|\delta|$ and the decay rate of transmon, we have the inverse of Hamiltonian of the quantum jump formalism under two driving fields $\hat{a}_1$ and $\hat{a}_2$
\begin{align} \nonumber
 \left( \hat{H}_{NH}^{(1)}\right)^{-1} = & (\Delta_a + \Delta_{in} - \Delta_{in})^{-1} |e\rangle \langle e| = |e\rangle \langle e| / \Delta_a\,,  \\ \nonumber
\left( \hat{H}_{NH}^{(2)}\right)^{-1} = & (\Delta_a + \Delta_{in} - \Delta_{in} - \delta)^{-1} |e\rangle \langle e| = |e\rangle \langle e| / (\Delta_a - \delta) \,.  \\ \nonumber
\end{align}
Thus, the effective Hamiltonian is given by
\begin{widetext}
\begin{equation}
 \begin{split}
  \hat{H}_{eff} & = -\frac{1}{2} \left [\hat{V}_- \sum_{j=1,2} \left( \hat{H}_{NH}^{(j)}\right)^{-1} \hat{V}_{+,j}\right ] + \hat{H}_g + \hat{H}_r \\
  & = -\frac{1}{2} \left[ \sum_{j=1,2} g_j^* \hat{a}_j^\dag \left( \frac{g_1 \hat{a}_1}{\Delta_a}  + \frac{g_2 \hat{a}_2}{\Delta_a - \delta}\right)|g\rangle \langle g| + H.c. \right] + \hat{H}_g + \hat{H}_r \\
  & = \hat{H}_g + \hat{H}_r- \frac{|g_1|^2}{\Delta_a} \hat{a}_1^\dag \hat{a}_1|g\rangle \langle g| - \frac{|g_2|^2}{\Delta_a - \delta} \hat{a}_2^\dag \hat{a}_2 |g\rangle \langle g| 
  \quad - \frac{1}{2} \left(\frac{1}{\Delta_a} + \frac{1}{\Delta_a - \delta}\right)  (g_1 g_2^* \hat{a}_1 \hat{a}_2^\dag+ H.c.)|g\rangle \langle g| \,. 
 \end{split}
\end{equation}
\end{widetext}
It is reasonable to assume $\hat{\sigma}_{ee} \sim 0$ and $\hat{\sigma}_{gg} \sim 1$ when $\hat{\tilde\sigma}_+$ varies slowly and the population in $|e\rangle$ is small in the situation of $|\Delta_a| \gg |g_1|,|g_2|$. For simplicity, we also assume the two CPW resonators are identical, $\delta=0$ and $g_1=g_2=g$. This gives the effective Hamiltonian in Eq. (\ref{eq:EffH}) in agreement with \cite{Raman2}.

\bibliographystyle{apsrev4-1}

\begin{thebibliography}{10}%
\makeatletter
\providecommand \@ifxundefined [1]{%
 \ifx #1\undefined \expandafter \@firstoftwo
 \else \expandafter \@secondoftwo
\fi
}%
\providecommand \@ifnum [1]{%
 \ifnum #1\expandafter \@firstoftwo
 \else \expandafter \@secondoftwo
\fi
}%
\providecommand \enquote [1]{``#1''}%
\providecommand \bibnamefont  [1]{#1}%
\providecommand \bibfnamefont [1]{#1}%
\providecommand \citenamefont [1]{#1}%
\providecommand\href[0]{\@sanitize\@href}%
\providecommand\@href[1]{\endgroup\@@startlink{#1}\endgroup\@@href}%
\providecommand\@@href[1]{#1\@@endlink}%
\providecommand \@sanitize [0]{\begingroup\catcode`\&12\catcode`\#12\relax}%
\@ifxundefined \pdfoutput {\@firstoftwo}{%
 \@ifnum{\z@=\pdfoutput}{\@firstoftwo}{\@secondoftwo}%
}{%
 \providecommand\@@startlink[1]{\leavevmode}%
 \providecommand\@@endlink[0]{}%
}{%
 \providecommand\@@startlink[1]{%
  \leavevmode
  \pdfstartlink
   attr{/Border[0 0 1 ]/H/I/C[0 1 1]}%
   user{/Subtype/Link/A<</Type/Action/S/URI/URI(#1)>>}%
  \relax
 }%
 \providecommand\@@endlink[0]{\pdfendlink}%
}%
\providecommand \url  [0]{\begingroup\@sanitize \@url }%
\providecommand \@url [1]{\endgroup\@href {#1}{\urlprefix}}%
\providecommand \urlprefix [0]{URL }%
\providecommand \Eprint[0]{\href }%
\@ifxundefined \urlstyle {%
  \providecommand \doi [1]{doi:\discretionary{}{}{}#1}%
}{%
  \providecommand \doi [0]{doi:\discretionary{}{}{}\begingroup
  \urlstyle{rm}\Url }%
}%
\providecommand \doibase [0]{http://dx.doi.org/}%
\providecommand \Doi[1]{\href{\doibase#1}}%
\providecommand \bibAnnote [3]{%
  \BibitemShut{#1}%
  \begin{quotation}\noindent
    \textsc{Key:}\ #2\\\textsc{Annotation:}\ #3%
  \end{quotation}%
}%
\providecommand \bibAnnoteFile [2]{%
  \IfFileExists{#2}{\bibAnnote {#1} {#2} {\input{#2}}}{}%
}%
\providecommand \typeout [0]{\immediate \write \m@ne }%
\providecommand \selectlanguage [0]{\@gobble}%
\providecommand \bibinfo [0]{\@secondoftwo}%
\providecommand \bibfield [0]{\@secondoftwo}%
\providecommand \translation [1]{[#1]}%
\providecommand \BibitemOpen[0]{}%
\providecommand \bibitemStop [0]{}%
\providecommand \bibitemNoStop [0]{.\EOS\space}%
\providecommand \EOS [0]{\spacefactor3000\relax}%
\providecommand \BibitemShut [1]{\csname bibitem#1\endcsname}%
\bibitem{AEIT1}%
  \BibitemOpen
  \bibfield{author}{%
  \bibinfo {author} {\bibfnamefont{S.~E.}\ \bibnamefont{Harris}}, \bibinfo
  {author} {\bibfnamefont{J.~E.}\ \bibnamefont{Field}},\ and\ \bibinfo {author}
  {\bibfnamefont{A.}~\bibnamefont{Imamo\v{g}lu}},\ }%
  \bibfield{journal}{%
  \bibinfo {journal} {Phys. Rev. Lett.}\ }%
  \textbf{\bibinfo {volume} {64}},\ \bibinfo {pages} {1107} (\bibinfo {year}
  {1990})%
  \bibAnnoteFile{NoStop}{AEIT1}%
\bibitem{AEIT2}%
  \BibitemOpen
  \bibfield{author}{%
  \bibinfo {author} {\bibfnamefont{M.}~\bibnamefont{Fleischhauer}}, \bibinfo
  {author} {\bibfnamefont{A.}~\bibnamefont{Imamo\v{g}lu}},\ and\ \bibinfo
  {author} {\bibfnamefont{J.~P.}\ \bibnamefont{Marangos}},\ }%
  \bibfield{journal}{%
  \bibinfo {journal} {Rev. Mod. Phys.}\ }%
  \textbf{\bibinfo {volume} {77}},\ \bibinfo {pages} {633} (\bibinfo {year}
  {2005})%
  \bibAnnoteFile{NoStop}{AEIT2}%
\bibitem{Kerr}%
  \BibitemOpen
  \bibfield{author}{%
  \bibinfo {author} {\bibfnamefont{Y.}~\bibnamefont{Niu}}, \bibinfo {author}
  {\bibfnamefont{S.}~\bibnamefont{Gong}}, \bibinfo {author}
  {\bibfnamefont{R.}~\bibnamefont{Li}}, \bibinfo {author}
  {\bibfnamefont{Z.}~\bibnamefont{Xu}},\ and\ \bibinfo {author}
  {\bibfnamefont{X.}~\bibnamefont{Liang}},\ }%
  \bibfield{journal}{%
  \bibinfo {journal} {Opt. Lett.}\ }%
  \textbf{\bibinfo {volume} {30}},\ \bibinfo {pages} {3371} (\bibinfo {year}
  {2005})%
  \bibAnnoteFile{NoStop}{Kerr}%
\bibitem{AStore1}%
  \BibitemOpen
  \bibfield{author}{%
  \bibinfo {author} {\bibfnamefont{M.~D.}\ \bibnamefont{Eisaman}}, \bibinfo
  {author} {\bibfnamefont{A.}~\bibnamefont{Andr\'{e}}}, \bibinfo {author}
  {\bibfnamefont{F.}~\bibnamefont{Massou}}, \bibinfo {author}
  {\bibfnamefont{M.}~\bibnamefont{Fleischhauer}}, \bibinfo {author}
  {\bibfnamefont{A.~S.}\ \bibnamefont{Zibrov}},\ and\ \bibinfo {author}
  {\bibfnamefont{M.~D.}\ \bibnamefont{Lukin}},\ }%
  \bibfield{journal}{%
  \bibinfo {journal} {Nature}\ }%
  \textbf{\bibinfo {volume} {438}},\ \bibinfo {pages} {837} (\bibinfo {year}
  {2005})%
  \bibAnnoteFile{NoStop}{AStore1}%
\bibitem{OEIT1}%
  \BibitemOpen
  \bibfield{author}{%
  \bibinfo {author} {\bibfnamefont{K.}~\bibnamefont{Totsuka}}, \bibinfo
  {author} {\bibfnamefont{N.}~\bibnamefont{Kobayashi}},\ and\ \bibinfo {author}
  {\bibfnamefont{M.}~\bibnamefont{Tomita}},\ }%
  \bibfield{journal}{%
  \bibinfo {journal} {Phys. Rev. Lett.}\ }%
  \textbf{\bibinfo {volume} {98}},\ \bibinfo {pages} {213904} (\bibinfo {year}
  {2007})%
  \bibAnnoteFile{NoStop}{OEIT1}%
\bibitem{OEIT2}%
  \BibitemOpen
  \bibfield{author}{%
  \bibinfo {author} {\bibfnamefont{Q.}~\bibnamefont{Xu}}, \bibinfo {author}
  {\bibfnamefont{P.}~\bibnamefont{Dong}},\ and\ \bibinfo {author}
  {\bibfnamefont{M.}~\bibnamefont{Lipson}},\ }%
  \bibfield{journal}{%
  \bibinfo {journal} {Nature Phys.}\ }%
  \textbf{\bibinfo {volume} {3}},\ \bibinfo {pages} {406} (\bibinfo {year}
  {2007})%
  \bibAnnoteFile{NoStop}{OEIT2}%
\bibitem{OEIT3}%
  \BibitemOpen
  \bibfield{author}{%
  \bibinfo {author} {\bibfnamefont{C.}~\bibnamefont{Zheng}}, \bibinfo {author}
  {\bibfnamefont{X.}~\bibnamefont{Jiang}}, \bibinfo {author}
  {\bibfnamefont{S.}~\bibnamefont{Hua}}, \bibinfo {author}
  {\bibfnamefont{L.}~\bibnamefont{Chang}}, \bibinfo {author}
  {\bibfnamefont{G.}~\bibnamefont{Li}}, \bibinfo {author}
  {\bibfnamefont{H.}~\bibnamefont{Fan}},\ and\ \bibinfo {author}
  {\bibfnamefont{M.}~\bibnamefont{Xiao}},\ }%
  \bibfield{journal}{%
  \bibinfo {journal} {Opt. Express}\ }%
  \textbf{\bibinfo {volume} {20}},\ \bibinfo {pages} {18319} (\bibinfo {year}
  {2012})%
  \bibAnnoteFile{NoStop}{OEIT3}%
\bibitem{OEIT4}%
  \BibitemOpen
  \bibfield{author}{%
  \bibinfo {author} {\bibfnamefont{X.}~\bibnamefont{Yang}}, \bibinfo {author}
  {\bibfnamefont{M.}~\bibnamefont{Yu}}, \bibinfo {author}
  {\bibfnamefont{D.}~\bibnamefont{Kwong}},\ and\ \bibinfo {author}
  {\bibfnamefont{C.~W.}\ \bibnamefont{Wong}},\ }%
  \bibfield{journal}{%
  \bibinfo {journal} {Phys. Rev. Lett.}\ }%
  \textbf{\bibinfo {volume} {102}},\ \bibinfo {pages} {173902} (\bibinfo {year}
  {2009})%
  \bibAnnoteFile{NoStop}{OEIT4}%
\bibitem{OCavity1}%
  \BibitemOpen
  \bibfield{author}{%
  \bibinfo {author} {\bibfnamefont{S.~I.}\ \bibnamefont{Schmid}}, \bibinfo
  {author} {\bibfnamefont{K.}~\bibnamefont{Xia}},\ and\ \bibinfo {author}
  {\bibfnamefont{J.}~\bibnamefont{Evers}},\ }%
  \bibfield{journal}{%
  \bibinfo {journal} {Phys. Rev. A}\ }%
  \textbf{\bibinfo {volume} {84}},\ \bibinfo {pages} {013808} (\bibinfo {year}
  {2011})%
  \bibAnnoteFile{NoStop}{OCavity1}%
\bibitem{OEMIT1}%
  \BibitemOpen
  \bibfield{author}{%
  \bibinfo {author} {\bibfnamefont{S.}~\bibnamefont{Weis}}, \bibinfo {author}
  {\bibfnamefont{R.}~\bibnamefont{Rivi\`{e}re}}, \bibinfo {author}
  {\bibfnamefont{S.}~\bibnamefont{Del\'{e}glise}}, \bibinfo {author}
  {\bibfnamefont{E.}~\bibnamefont{Gavartin}}, \bibinfo {author}
  {\bibfnamefont{O.}~\bibnamefont{Arcizet}}, \bibinfo {author}
  {\bibfnamefont{A.}~\bibnamefont{Schliesser}},\ and\ \bibinfo {author}
  {\bibfnamefont{T.~J.}\ \bibnamefont{Kippenberg}},\ }%
  \bibfield{journal}{%
  \bibinfo {journal} {Science}\ }%
  \textbf{\bibinfo {volume} {330}},\ \bibinfo {pages} {1520} (\bibinfo {year}
  {2010})%
  \bibAnnoteFile{NoStop}{OEMIT1}%
\bibitem{OEMIT2}%
  \BibitemOpen
  \bibfield{author}{%
  \bibinfo {author} {\bibfnamefont{A.~H.}\ \bibnamefont{Safavi-Naeini}},
  \bibinfo {author} {\bibfnamefont{T.~P.}\ \bibnamefont{{Mayer Alegre}}},
  \bibinfo {author} {\bibfnamefont{J.}~\bibnamefont{Chan}}, \bibinfo {author}
  {\bibfnamefont{M.}~\bibnamefont{Eichenfield}}, \bibinfo {author}
  {\bibfnamefont{M.}~\bibnamefont{Winger}}, \bibinfo {author}
  {\bibfnamefont{Q.}~\bibnamefont{Lin}}, \bibinfo {author}
  {\bibfnamefont{J.~T.}\ \bibnamefont{Hill}}, \bibinfo {author}
  {\bibfnamefont{D.~E.}\ \bibnamefont{Chang}},\ and\ \bibinfo {author}
  {\bibfnamefont{O.}~\bibnamefont{Painter}},\ }%
  \bibfield{journal}{%
  \bibinfo {journal} {Nature}\ }%
  \textbf{\bibinfo {volume} {472}},\ \bibinfo {pages} {69} (\bibinfo {year}
  {2011})%
  \bibAnnoteFile{NoStop}{OEMIT2}%
\bibitem{MWPhoton}%
  \BibitemOpen
  \bibfield{author}{%
  \bibinfo {author} {\bibfnamefont{J.}~\bibnamefont{Capmany}}, \bibinfo
  {author} {\bibfnamefont{I.}~\bibnamefont{Gasulla}},\ and\ \bibinfo {author}
  {\bibfnamefont{S.}~\bibnamefont{Sales}},\ }%
  \bibfield{journal}{%
  \bibinfo {journal} {Nature Photon.}\ }%
  \textbf{\bibinfo {volume} {5}},\ \bibinfo {pages} {731} (\bibinfo {year}
  {2011})%
  \bibAnnoteFile{NoStop}{MWPhoton}%
\bibitem{MWDelay3}%
  \BibitemOpen
  \bibfield{author}{%
  \bibinfo {author} {\bibfnamefont{N.}~\bibnamefont{Papasimakis}}, \bibinfo
  {author} {\bibfnamefont{V.~A.}\ \bibnamefont{Fedotov}}, \bibinfo {author}
  {\bibfnamefont{N.~I.}\ \bibnamefont{Zheludev}},\ and\ \bibinfo {author}
  {\bibfnamefont{S.~L.}\ \bibnamefont{Prosvirnin}},\ }%
  \bibfield{journal}{%
  \bibinfo {journal} {Phys. Rev. Lett.}\ }%
  \textbf{\bibinfo {volume} {101}},\ \bibinfo {pages} {253903} (\bibinfo {year}
  {2008})%
  \bibAnnoteFile{NoStop}{MWDelay3}%
\bibitem{MWDelay1}%
  \BibitemOpen
  \bibfield{author}{%
  \bibinfo {author} {\bibfnamefont{C.}~\bibnamefont{Jiang}}, \bibinfo {author}
  {\bibfnamefont{B.}~\bibnamefont{Chen}},\ and\ \bibinfo {author}
  {\bibfnamefont{K.}~\bibnamefont{Zhu}},\ }%
  \bibfield{journal}{%
  \bibinfo {journal} {Europhys. Lett.}\ }%
  \textbf{\bibinfo {volume} {94}},\ \bibinfo {pages} {38002} (\bibinfo {year}
  {2011})%
  \bibAnnoteFile{NoStop}{MWDelay1}%
\bibitem{MWDelay2}%
  \BibitemOpen
  \bibfield{author}{%
  \bibinfo {author} {\bibfnamefont{X.}~\bibnamefont{Zhou}}, \bibinfo {author}
  {\bibfnamefont{F.}~\bibnamefont{Hocke}}, \bibinfo {author}
  {\bibfnamefont{A.}~\bibnamefont{Schliesser}}, \bibinfo {author}
  {\bibfnamefont{A.}~\bibnamefont{Marx}}, \bibinfo {author}
  {\bibfnamefont{H.}~\bibnamefont{Huebl}}, \bibinfo {author}
  {\bibfnamefont{R.}~\bibnamefont{Gross}},\ and\ \bibinfo {author}
  {\bibfnamefont{T.~J.}\ \bibnamefont{Kippenberg}},\ }%
  \bibfield{journal}{%
  \bibinfo {journal} {Nature Phys.}\ }%
  \textbf{\bibinfo {volume} {9}},\ \bibinfo {pages} {179} (\bibinfo {year}
  {2013})%
  \bibAnnoteFile{NoStop}{MWDelay2}%
\bibitem{StoreMWinEIT}%
  \BibitemOpen
  \bibfield{author}{%
  \bibinfo {author} {\bibfnamefont{T.}~\bibnamefont{Nakanishi}}, \bibinfo
  {author} {\bibfnamefont{T.}~\bibnamefont{Otani}}, \bibinfo {author}
  {\bibfnamefont{Y.}~\bibnamefont{Tamayama}},\ and\ \bibinfo {author}
  {\bibfnamefont{M.}~\bibnamefont{Kitano}},\ }%
  \bibfield{journal}{%
  \bibinfo {journal} {Phys. Rev. B}\ }%
  \textbf{\bibinfo {volume} {87}},\ \bibinfo {pages} {161110} (\bibinfo {year}
  {2013})%
  \bibAnnoteFile{NoStop}{StoreMWinEIT}%
\bibitem{MWStorage1}%
  \BibitemOpen
  \bibfield{author}{%
  \bibinfo {author} {\bibfnamefont{H.}~\bibnamefont{Wu}}, \bibinfo {author}
  {\bibfnamefont{R.~E.}\ \bibnamefont{George}}, \bibinfo {author}
  {\bibfnamefont{J.~H.}\ \bibnamefont{Wesenberg}}, \bibinfo {author}
  {\bibfnamefont{K.}~\bibnamefont{M\o{}lmer}}, \bibinfo {author}
  {\bibfnamefont{D.~I.}\ \bibnamefont{Schuster}}, \bibinfo {author}
  {\bibfnamefont{R.~J.}\ \bibnamefont{Schoelkopf}}, \bibinfo {author}
  {\bibfnamefont{K.~M.}\ \bibnamefont{Itoh}}, \bibinfo {author}
  {\bibfnamefont{A.}~\bibnamefont{Ardavan}}, \bibinfo {author}
  {\bibfnamefont{J.~J.~L.}\ \bibnamefont{Morton}},\ and\ \bibinfo {author}
  {\bibfnamefont{G.~A.~D.}\ \bibnamefont{Briggs}},\ }%
  \bibfield{journal}{%
  \bibinfo {journal} {Phys. Rev. Lett.}\ }%
  \textbf{\bibinfo {volume} {105}},\ \bibinfo {pages} {140503} (\bibinfo {year}
  {2010})%
  \bibAnnoteFile{NoStop}{MWStorage1}%
\bibitem{QMemory}%
  \BibitemOpen
  \bibfield{author}{%
  \bibinfo {author} {\bibfnamefont{B.}~\bibnamefont{Julsgaard}}, \bibinfo
  {author} {\bibfnamefont{C.}~\bibnamefont{Grezes}}, \bibinfo {author}
  {\bibfnamefont{P.}~\bibnamefont{Bertet}},\ and\ \bibinfo {author}
  {\bibfnamefont{K.}~\bibnamefont{M\o{}lmer}},\ }%
  \bibfield{journal}{%
  \bibinfo {journal} {Phys. Rev. Lett.}\ }%
  \textbf{\bibinfo {volume} {110}},\ \bibinfo {pages} {250503} (\bibinfo {year}
  {2013})%
  \bibAnnoteFile{NoStop}{QMemory}%
\bibitem{MWTransfer}%
  \BibitemOpen
  \bibfield{author}{%
  \bibinfo {author} {\bibfnamefont{T.~A.}\ \bibnamefont{Palomaki}}, \bibinfo
  {author} {\bibfnamefont{J.~W.}\ \bibnamefont{Harlow}}, \bibinfo {author}
  {\bibfnamefont{J.~D.}\ \bibnamefont{Teufel}}, \bibinfo {author}
  {\bibfnamefont{R.~W.}\ \bibnamefont{Simmonds}},\ and\ \bibinfo {author}
  {\bibfnamefont{K.~W.}\ \bibnamefont{Lehnert}},\ }%
  \bibfield{journal}{%
  \bibinfo {journal} {Nature}\ }%
  \textbf{\bibinfo {volume} {495}},\ \bibinfo {pages} {201} (\bibinfo {year}
  {2013})%
  \bibAnnoteFile{NoStop}{MWTransfer}%
\bibitem{MWStorage2}%
  \BibitemOpen
  \bibfield{author}{%
  \bibinfo {author} {\bibfnamefont{Y.}~\bibnamefont{Yin}}, \bibinfo {author}
  {\bibfnamefont{Y.}~\bibnamefont{Chen}}, \bibinfo {author}
  {\bibfnamefont{D.}~\bibnamefont{Sank}}, \bibinfo {author}
  {\bibfnamefont{P.~J.~J.}\ \bibnamefont{O'Malley}}, \bibinfo {author}
  {\bibfnamefont{T.~C.}\ \bibnamefont{White}}, \bibinfo {author}
  {\bibfnamefont{R.}~\bibnamefont{Barends}}, \bibinfo {author}
  {\bibfnamefont{J.}~\bibnamefont{Kelly}}, \bibinfo {author}
  {\bibfnamefont{E.}~\bibnamefont{Lucero}}, \bibinfo {author}
  {\bibfnamefont{M.}~\bibnamefont{Mariantoni}}, \bibinfo {author}
  {\bibfnamefont{A.}~\bibnamefont{Megrant}}, \bibinfo {author}
  {\bibfnamefont{C.}~\bibnamefont{Neill}}, \bibinfo {author}
  {\bibfnamefont{A.}~\bibnamefont{Vainsencher}}, \bibinfo {author}
  {\bibfnamefont{J.}~\bibnamefont{Wenner}}, \bibinfo {author}
  {\bibfnamefont{A.~N.}\ \bibnamefont{Korotkov}}, \bibinfo {author}
  {\bibfnamefont{A.~N.}\ \bibnamefont{Cleland}},\ and\ \bibinfo {author}
  {\bibfnamefont{J.~M.}\ \bibnamefont{Martinis}},\ }%
  \bibfield{journal}{%
  \bibinfo {journal} {Phys. Rev. Lett.}\ }%
  \textbf{\bibinfo {volume} {110}},\ \bibinfo {pages} {107001} (\bibinfo {year}
  {2013})%
  \bibAnnoteFile{NoStop}{MWStorage2}%
\bibitem{AT1}%
  \BibitemOpen
  \bibfield{author}{%
  \bibinfo {author} {\bibfnamefont{M.~A.}\ \bibnamefont{Sillanp\"{a}\"{a}}},
  \bibinfo {author} {\bibfnamefont{J.}~\bibnamefont{Li}}, \bibinfo {author}
  {\bibfnamefont{K.}~\bibnamefont{Cicak}}, \bibinfo {author}
  {\bibfnamefont{F.}~\bibnamefont{Altomare}}, \bibinfo {author}
  {\bibfnamefont{J.~I.}\ \bibnamefont{Park}}, \bibinfo {author}
  {\bibfnamefont{R.~W.}\ \bibnamefont{Simmonds}}, \bibinfo {author}
  {\bibfnamefont{G.~S.}\ \bibnamefont{Paraoanu}},\ and\ \bibinfo {author}
  {\bibfnamefont{P.~J.}\ \bibnamefont{Hakonen}},\ }%
  \bibfield{journal}{%
  \bibinfo {journal} {Phys. Rev. Lett.}\ }%
  \textbf{\bibinfo {volume} {103}},\ \bibinfo {pages} {193601} (\bibinfo {year}
  {2009})%
  \bibAnnoteFile{NoStop}{AT1}%
\bibitem{AT2}%
  \BibitemOpen
  \bibfield{author}{%
  \bibinfo {author} {\bibfnamefont{J.}~\bibnamefont{Li}}, \bibinfo {author}
  {\bibfnamefont{G.~S.}\ \bibnamefont{Paraoanu}}, \bibinfo {author}
  {\bibfnamefont{K.}~\bibnamefont{Cicak}}, \bibinfo {author}
  {\bibfnamefont{F.}~\bibnamefont{Altomare}}, \bibinfo {author}
  {\bibfnamefont{J.~I.}\ \bibnamefont{Park}}, \bibinfo {author}
  {\bibfnamefont{R.~W.}\ \bibnamefont{Simmonds}}, \bibinfo {author}
  {\bibfnamefont{M.~A.}\ \bibnamefont{Sillanp\"{a}\"{a}}},\ and\ \bibinfo
  {author} {\bibfnamefont{P.~J.}\ \bibnamefont{Hakonen}},\ }%
  \bibfield{journal}{%
  \bibinfo {journal} {Phys. Rev. B}\ }%
  \textbf{\bibinfo {volume} {84}},\ \bibinfo {pages} {104527} (\bibinfo {year}
  {2011})%
  \bibAnnoteFile{NoStop}{AT2}%
\bibitem{AT3}%
  \BibitemOpen
  \bibfield{author}{%
  \bibinfo {author} {\bibfnamefont{J.}~\bibnamefont{Li}}, \bibinfo {author}
  {\bibfnamefont{G.~S.}\ \bibnamefont{Paraoanu}}, \bibinfo {author}
  {\bibfnamefont{K.}~\bibnamefont{Cicak}}, \bibinfo {author}
  {\bibfnamefont{F.}~\bibnamefont{Altomare}}, \bibinfo {author}
  {\bibfnamefont{J.~I.}\ \bibnamefont{Park}}, \bibinfo {author}
  {\bibfnamefont{R.~W.}\ \bibnamefont{Simmonds}}, \bibinfo {author}
  {\bibfnamefont{M.~A.}\ \bibnamefont{Sillanp\"{a}\"{a}}},\ and\ \bibinfo
  {author} {\bibfnamefont{P.~J.}\ \bibnamefont{Hakonen}},\ }%
  \bibfield{journal}{%
  \bibinfo {journal} {Sci. Rep.}}%
  \ \doi{10.1038},
   (\bibinfo {year} {2012})%
  \bibAnnoteFile{NoStop}{AT3}%
\bibitem{Idea}%
  \BibitemOpen
  \bibfield{author}{%
  \bibinfo {author} {\bibfnamefont{K.}~\bibnamefont{Xia}}\ and\ \bibinfo
  {author} {\bibfnamefont{J.}~\bibnamefont{Twamley}},\ }%
  \bibfield{journal}{%
  \bibinfo {journal} {Phys. Rev. X}\ }%
  \textbf{\bibinfo {volume} {3}},\ \bibinfo {pages} {031013} (\bibinfo {year}
  {2013})%
  \bibAnnoteFile{NoStop}{Idea}%
\bibitem{Design}%
  \BibitemOpen
  \bibfield{author}{%
  \bibinfo {author} {\bibfnamefont{D.~I.}\ \bibnamefont{Schuster}}, \bibinfo
  {author} {\bibfnamefont{A.~A.}\ \bibnamefont{Houck}}, \bibinfo {author}
  {\bibfnamefont{J.~A.}\ \bibnamefont{Schreier}}, \bibinfo {author}
  {\bibfnamefont{A.}~\bibnamefont{Wallraff}}, \bibinfo {author}
  {\bibfnamefont{J.~M.}\ \bibnamefont{Gambetta}}, \bibinfo {author}
  {\bibfnamefont{A.}~\bibnamefont{Blais}}, \bibinfo {author}
  {\bibfnamefont{L.}~\bibnamefont{Frunzio}}, \bibinfo {author}
  {\bibfnamefont{J.}~\bibnamefont{Majer}}, \bibinfo {author}
  {\bibfnamefont{B.}~\bibnamefont{Johnson}}, \bibinfo {author}
  {\bibfnamefont{M.~H.}\ \bibnamefont{Devoret}}, \bibinfo {author}
  {\bibfnamefont{S.~M.}\ \bibnamefont{Girvin}},\ and\ \bibinfo {author}
  {\bibfnamefont{R.~J.}\ \bibnamefont{Schoelkopf}},\ }%
  \bibfield{journal}{%
  \bibinfo {journal} {Nature}\ }%
  \textbf{\bibinfo {volume} {445}},\ \bibinfo {pages} {515} (\bibinfo {year}
  {2007})%
  \bibAnnoteFile{NoStop}{Design}%
\bibitem{ExpSetup}%
  \BibitemOpen
  \bibfield{author}{%
  \bibinfo {author} {\bibfnamefont{B.~R.}\ \bibnamefont{Johnson}}, \bibinfo
  {author} {\bibfnamefont{M.~D.}\ \bibnamefont{Reed}}, \bibinfo {author}
  {\bibfnamefont{A.~A.}\ \bibnamefont{Houck}}, \bibinfo {author}
  {\bibfnamefont{D.~I.}\ \bibnamefont{Schuster}}, \bibinfo {author}
  {\bibfnamefont{L.~S.}\ \bibnamefont{Bishop}}, \bibinfo {author}
  {\bibfnamefont{E.}~\bibnamefont{Ginossar}}, \bibinfo {author}
  {\bibfnamefont{J.~M.}\ \bibnamefont{Gambetta}}, \bibinfo {author}
  {\bibfnamefont{L.}~\bibnamefont{DiCarlo}}, \bibinfo {author}
  {\bibfnamefont{L.}~\bibnamefont{Frunzio}}, \bibinfo {author}
  {\bibfnamefont{S.~M.}\ \bibnamefont{Girvin}},\ and\ \bibinfo {author}
  {\bibfnamefont{R.~J.}\ \bibnamefont{Schoelkopf}},\ }%
  \bibfield{journal}{%
  \bibinfo {journal} {Nat. Phys.}\ }%
  \textbf{\bibinfo {volume} {6}},\ \bibinfo {pages} {663} (\bibinfo {year}
  {2010})%
  \bibAnnoteFile{NoStop}{ExpSetup}%
\bibitem{TransmonFluxBias}%
  \BibitemOpen
  \bibfield{author}{%
  \bibinfo {author} {\bibfnamefont{L.}~\bibnamefont{DiCarlo}}, \bibinfo
  {author} {\bibfnamefont{J.~M.}\ \bibnamefont{Chow}}, \bibinfo {author}
  {\bibfnamefont{J.~M.}\ \bibnamefont{Gambetta}}, \bibinfo {author}
  {\bibfnamefont{L.~S.}\ \bibnamefont{Bishop}}, \bibinfo {author}
  {\bibfnamefont{B.~R.}\ \bibnamefont{Johnson}}, \bibinfo {author}
  {\bibfnamefont{D.~I.}\ \bibnamefont{Schuster}}, \bibinfo {author}
  {\bibfnamefont{J.}~\bibnamefont{Majer}}, \bibinfo {author}
  {\bibfnamefont{A.}~\bibnamefont{Blais}}, \bibinfo {author}
  {\bibfnamefont{L.}~\bibnamefont{Frunzio}}, \bibinfo {author}
  {\bibfnamefont{S.~M.}\ \bibnamefont{Girvin}},\ and\ \bibinfo {author}
  {\bibfnamefont{R.~J.}\ \bibnamefont{Schoelkopf}},\ }%
  \bibfield{journal}{%
  \bibinfo {journal} {Nature}\ }%
  \textbf{\bibinfo {volume} {460}},\ \bibinfo {pages} {240} (\bibinfo {year}
  {2009})%
  \bibAnnoteFile{NoStop}{TransmonFluxBias}%
\bibitem{Tunefr1}%
  \BibitemOpen
  \bibfield{author}{%
  \bibinfo {author} {\bibfnamefont{A.}~\bibnamefont{Wallraff}}, \bibinfo
  {author} {\bibfnamefont{D.~I.}\ \bibnamefont{Schuster}}, \bibinfo {author}
  {\bibfnamefont{A.}~\bibnamefont{Blais}}, \bibinfo {author}
  {\bibfnamefont{L.}~\bibnamefont{Frunzio}}, \bibinfo {author}
  {\bibfnamefont{R.}~\bibnamefont{s.~Huang}}, \bibinfo {author}
  {\bibfnamefont{J.}~\bibnamefont{Majer}}, \bibinfo {author}
  {\bibfnamefont{S.}~\bibnamefont{Kumar}},\ and\ \bibinfo {author}
  {\bibfnamefont{S.~M.~G.}\ \bibnamefont{an~dR. J.~Schoelkopf}},\ }%
  \bibfield{journal}{%
  \bibinfo {journal} {Nature}\ }%
  \textbf{\bibinfo {volume} {431}},\ \bibinfo {pages} {162} (\bibinfo {year}
  {2004})%
  \bibAnnoteFile{NoStop}{Tunefr1}%
\bibitem{Tunefr2}%
  \BibitemOpen
  \bibfield{author}{%
  \bibinfo {author} {\bibfnamefont{Z.~L.}\ \bibnamefont{Wang}}, \bibinfo
  {author} {\bibfnamefont{Y.~P.}\ \bibnamefont{Zhong}}, \bibinfo {author}
  {\bibfnamefont{L.~J.}\ \bibnamefont{He}}, \bibinfo {author}
  {\bibfnamefont{H.}~\bibnamefont{Wang}}, \bibinfo {author}
  {\bibfnamefont{J.~M.}\ \bibnamefont{Martinis}},\ and\ \bibinfo {author}
  {\bibfnamefont{A.~N.}\ \bibnamefont{Cleland}},\ }%
  \bibfield{journal}{%
  \bibinfo {journal} {Appl. Phys. Lett.}\ }%
  \textbf{\bibinfo {volume} {102}},\ \bibinfo {pages} {163503} (\bibinfo {year}
  {2013})%
  \bibAnnoteFile{NoStop}{Tunefr2}%
\bibitem{Tunefr3}%
  \BibitemOpen
  \bibfield{author}{%
  \bibinfo {author} {\bibfnamefont{A.}~\bibnamefont{Palacios-Laloy}}, \bibinfo
  {author} {\bibfnamefont{F.}~\bibnamefont{Nguyen}}, \bibinfo {author}
  {\bibfnamefont{F.}~\bibnamefont{Mallet}}, \bibinfo {author}
  {\bibfnamefont{P.}~\bibnamefont{Bertet}}, \bibinfo {author}
  {\bibfnamefont{D.}~\bibnamefont{Vion}},\ and\ \bibinfo {author}
  {\bibfnamefont{D.}~\bibnamefont{Esteve}},\ }%
  \bibfield{journal}{%
  \bibinfo {journal} {J. Low Temp. Phys.}\ }%
  \textbf{\bibinfo {volume} {151}},\ \bibinfo {pages} {1034} (\bibinfo {year}
  {2008})%
  \bibAnnoteFile{NoStop}{Tunefr3}%
\bibitem{InputOutput1}%
  \BibitemOpen
  \bibfield{author}{%
  \bibinfo {author} {\bibfnamefont{M.~J.}\ \bibnamefont{Collett}}\ and\
  \bibinfo {author} {\bibfnamefont{C.~W.}\ \bibnamefont{Gardiner}},\ }%
  \bibfield{journal}{%
  \bibinfo {journal} {Phys. Rev. A}\ }%
  \textbf{\bibinfo {volume} {30}},\ \bibinfo {pages} {1386} (\bibinfo {year}
  {1984})%
  \bibAnnoteFile{NoStop}{InputOutput1}%
\bibitem{InputOutput2}%
  \BibitemOpen
  \bibfield{author}{%
  \bibinfo {author} {\bibfnamefont{C.~W.}\ \bibnamefont{Gardiner}}\ and\
  \bibinfo {author} {\bibfnamefont{M.~J.}\ \bibnamefont{Collett}},\ }%
  \bibfield{journal}{%
  \bibinfo {journal} {Phys. Rev. A}\ }%
  \textbf{\bibinfo {volume} {31}},\ \bibinfo {pages} {3761} (\bibinfo {year}
  {1985})%
  \bibAnnoteFile{NoStop}{InputOutput2}%
\bibitem{OL30p2001}%
  \BibitemOpen
  \bibfield{author}{%
  \bibinfo {author} {\bibfnamefont{J.~T.}\ \bibnamefont{Shen}}\ and\ \bibinfo
  {author} {\bibfnamefont{S.}~\bibnamefont{Fan}},\ }%
  \bibfield{journal}{%
  \bibinfo {journal} {Opt. Lett.}\ }%
  \textbf{\bibinfo {volume} {30}},\ \bibinfo {pages} {2001} (\bibinfo {year}
  {2005})%
  \bibAnnoteFile{NoStop}{OL30p2001}%
\bibitem{PRA79p023837}%
  \BibitemOpen
  \bibfield{author}{%
  \bibinfo {author} {\bibfnamefont{J.}~\bibnamefont{Shen}}\ and\ \bibinfo
  {author} {\bibfnamefont{S.}~\bibnamefont{Fan}},\ }%
  \bibfield{journal}{%
  \bibinfo {journal} {Phys. Rev. A}\ }%
  \textbf{\bibinfo {volume} {79}},\ \bibinfo {pages} {023837} (\bibinfo {year}
  {2009})%
  \bibAnnoteFile{NoStop}{PRA79p023837}%
\bibitem{PRA79p023838}%
  \BibitemOpen
  \bibfield{author}{%
  \bibinfo {author} {\bibfnamefont{J.}~\bibnamefont{Shen}}\ and\ \bibinfo
  {author} {\bibfnamefont{S.}~\bibnamefont{Fan}},\ }%
  \bibfield{journal}{%
  \bibinfo {journal} {Phys. Rev. A}\ }%
  \textbf{\bibinfo {volume} {79}},\ \bibinfo {pages} {023838} (\bibinfo {year}
  {2009})%
  \bibAnnoteFile{NoStop}{PRA79p023838}%
\bibitem{PRA82p063839}%
  \BibitemOpen
  \bibfield{author}{%
  \bibinfo {author} {\bibfnamefont{E.~E.~H.}\ \bibnamefont{III}}, \bibinfo
  {author} {\bibfnamefont{A.~W.}\ \bibnamefont{Elshaari}},\ and\ \bibinfo
  {author} {\bibfnamefont{S.~F.}\ \bibnamefont{Preble}},\ }%
  \bibfield{journal}{%
  \bibinfo {journal} {Phys. Rev. A}\ }%
  \textbf{\bibinfo {volume} {82}},\ \bibinfo {pages} {063839} (\bibinfo {year}
  {2010})%
  \bibAnnoteFile{NoStop}{PRA82p063839}%
\bibitem{Raman1}%
  \BibitemOpen
  \bibfield{author}{%
  \bibinfo {author} {\bibfnamefont{F.}~\bibnamefont{Reiter}}\ and\ \bibinfo
  {author} {\bibfnamefont{A.~S.}\ \bibnamefont{S\o{}rensen}},\ }%
  \bibfield{journal}{%
  \bibinfo {journal} {Phys. Rev. A}\ }%
  \textbf{\bibinfo {volume} {85}},\ \bibinfo {pages} {032111} (\bibinfo {year}
  {2012})%
  \bibAnnoteFile{NoStop}{Raman1}%
\bibitem{Raman2}%
  \BibitemOpen
  \bibfield{author}{%
  \bibinfo {author} {\bibfnamefont{J.~I.}\ \bibnamefont{Cirac}}, \bibinfo
  {author} {\bibfnamefont{P.}~\bibnamefont{Zoller}}, \bibinfo {author}
  {\bibfnamefont{H.~J.}\ \bibnamefont{Kimble}},\ and\ \bibinfo {author}
  {\bibfnamefont{H.}~\bibnamefont{Mabuchi}},\ }%
  \bibfield{journal}{%
  \bibinfo {journal} {Phys. Rev. Lett.}\ }%
  \textbf{\bibinfo {volume} {78}},\ \bibinfo {pages} {3221} (\bibinfo {year}
  {1997})%
  \bibAnnoteFile{NoStop}{Raman2}%
\end{thebibliography}

%

\end{document}